\documentclass[11pt,a4paper]{article}
\pdfoutput=1
\usepackage{graphicx}
\usepackage{jcappub}
\usepackage{amssymb}
\usepackage{amsmath,amssymb,amsfonts}
\usepackage{comment}
\usepackage{hyperref}
\usepackage{xcolor}
\usepackage{multicol}
\usepackage{placeins}

\newcommand{\be}{\begin{equation}}
\newcommand{\ee}{\end{equation}}
\newcommand{\bea}{\begin{eqnarray}}
\newcommand{\eea}{\end{eqnarray}}

\makeindex

\begin{document}
\hypersetup{pageanchor=false} 

\title{Bayesian evidence and predictivity of the inflationary paradigm}

\author[a]{Giulia Gubitosi}
\author[a,b]{Macarena Lagos}
\author[a]{Jo\~{a}o Magueijo}
\author[b]{Rupert Allison}

\affiliation[a]{Theoretical Physics, Blackett Laboratory, Imperial College, London, SW7 2BZ,
UK}
\affiliation[b]{Astrophysics, University of Oxford, DWB, Keble Road, Oxford OX1 3RH, UK}

\emailAdd{g.gubitosi@imperial.ac.uk}
\emailAdd{m.lagos13@imperial.ac.uk}
\emailAdd{j.magueijo@imperial.ac.uk}
\emailAdd{rupert.allison@astro.ox.ac.uk}

\abstract{In this paper we consider the issue of paradigm evaluation by applying Bayes' theorem along the following nested hierarchy of progressively more complex structures: i) parameter estimation (within a model), ii) model selection and comparison (within a paradigm), iii) paradigm evaluation. In such a hierarchy the Bayesian evidence works both as the posterior's normalization at a given level and as the likelihood function at the next level up. Whilst raising no objections to the standard application of the procedure at the two lowest levels, we argue that it should receive a considerable modification when evaluating paradigms, when {\it testability} and fitting data are equally important. By considering toy models we illustrate how models and paradigms that are difficult to falsify are always favoured by the Bayes factor. We argue that the evidence for a paradigm should not only be high for a given dataset, but {\it exceptional} with respect to what it would have been, had the data been different. With this motivation we propose a measure which we term {\it predictivity}, as well as a prior to be incorporated into the Bayesian framework, penalising unpredictivity as much as not fitting data. We apply this measure to inflation seen as a whole, and to a scenario where a specific inflationary model is hypothetically deemed as the only one viable as a result of information alien to cosmology (e.g.~Solar System gravity experiments, or particle physics input). We conclude that cosmic inflation is currently hard to falsify, 
but that this could change were external/additional information to cosmology to select one of its many models. We also compare this state of affairs to bimetric varying speed of light cosmology.}


\maketitle


\section{Introduction}

As the cosmological data continues to improve with its inevitable twists, it has become evident that whatever the observations turn out to be they will be lauded as ``proof of inflation''. This was poignantly brought to the fore when the BICEP2 data was released \cite{Ade:2014xna}, in the wake of Planck's initial cosmological papers \cite{PlanckXVI,Ade:2013kta,Planck:2013jfk,Ade:2013ydc}. Even though the two datasets taken at face-value contradicted each other, they were both advertised as proof of inflation. With the demise of the BICEP2 claim no one seems to have noted the flaw subjacent to this attitude: inflation can in fact predict a large range of values for observables. Interesting sociology will no doubt be reenacted when Planck's polarisation data makes its mark, in the hopefully not too distant future. 

Independently of where the correct observations end up settling, matters such as model selection and paradigm evaluation 
will have to be addressed before any proper scientific conclusions are drawn. We should quantify the fact that if inflation seen as a whole (i.e.~as a {\it paradigm}) were indeed to ``fit anything'', then, reciprocally, it would not be possible to disprove it. 
Penalisation schemes for models which fit the data due to an abundance of free parameters do exist, but transposing these schemes to paradigms is far from obvious. In this paper we address this problem.

The Bayesian framework has been the preferred tool for assessing the performance of different models in cosmology and astrophysics \cite{Trotta:2008qt}. In its standard use, the Bayes factor is employed to compare models or for finding the best-fit parameters for a specific model. This Bayes factor includes 
an automatic Occam's razor that penalises unpredictive models (with large number or range of parameters), and thus it considers them to be less likely. However, it penalises far more harshly for not fitting well data. If we insist on testability being equally important as fitting data, the use of the Bayes factor is still perfectly adequate for the standard use because the various options under comparison are usually roughly equally predictive. However, limitations will arise when the options have very different predictive powers, which will be the case when comparing paradigms. 

In Section~\ref{3levels}, we show that the Bayes' theorem may be used in nested chains involving qualitatively very different levels, such as parameter estimation (within a given model), model selection (within a paradigm), and paradigm evaluation. This last chain on paradigm evaluation is used in the following sections to assess the performance of the inflationary paradigm (thought as the collection of inflationary models).

In Section~\ref{toy}, by means of a number of toy models, we show that the Bayes factor is more suited to playing the lottery than to paradigm testing, since it regards winning as more important than being predictive. In fact, missing the jackpot is penalised exponentially by the Bayes factor, whereas being unpredictive is only penalised as a power-law. We argue that a modification of the standard Bayesian calculations is required for paradigm testing, 
given that a successful paradigm should not only fit well the data, but do so
{\it having taken the full brunt of the risk of failure}. 
In Section~\ref{pred-measure} we use these toy models to motivate a measure of predictivity which can be incorporated into the Bayesian framework when testing paradigms. Using this definition of predictivity, in Section~\ref{new-factor} we construct a penalty factor which fines theories for being unpredictive as much as the Bayes factor penalises them for not fitting data. We present specific examples illustrating the issues at stake. 

The rest of the paper is devoted to the application of the formalism to a collection of 85 single-field inflationary models, selected for numerical convenience. In Section~\ref{PredInflationModels} we examine the evidences and predictivities of these models, confirming that at the level of model selection there is little point in correcting the Bayes factor. In Section \ref{SecPredInflation} we compute an upper bound for the predictivity of the inflationary paradigm, to find a rather low value. However, this would change dramatically should information external to cosmology become available, something we illustrate with an hypothetical situation where Higgs inflation is deemed {\it a priori} as the only viable inflationary model. 

The situation is somewhat different with alternative paradigms, such as the cyclic Universe and bimetric varying speed of light, as explained in Section~\ref{VSL}. In Section \ref{SecConclusions}, we summarise our findings and comment on their meaning within the bigger picture.

\section{Parameters, models and paradigms}\label{3levels}
At its most schematic Bayes' theorem reads:
\be
P(T|D)=\frac{P(D|T)\pi(T)}{P(D)},
\ee
where $D$ denotes ``data'' and $T$ ``theory'', with $T$ often standing for a variable (discrete or continuous) labelling a class of theories. To establish terminology we recall that $P(D|T)$ is also referred to as the ``likelihood'' of the theory (denoted ${\cal L}(T)$), $\pi(T)$ as its ``prior'' (i.e.~the prior probability of the theory before data $D$ is considered), and $P(T|D)$ as its ``posterior'' (i.e.~the posterior probability of the theory after the data was considered). The denominator $P(D)$ normalises the posterior distribution and can be obtained from 
\be
P(D)=\int dT\, P(D|T)\pi(T).
\ee
This is also the probability of the data given the whole class of theories $\{T\}$, and so forms the Bayesian evidence
for this class:
\be
{\cal E}=P(D).
\ee

Bayes' theorem is frequently applied in chains; for example as new datasets are considered we may recycle the posterior resulting from previous datasets into the prior to be used in the analysis of the new dataset. Here we shall consider a different type of chain, resulting from assessing the effects of the {\it same} dataset $D$ at the three successive, qualitatively different levels:
\begin{itemize}
\item Level 1. Finding the best {\bf parameters} for a given model.
\item Level 2. Comparing the relative merits of different {\bf models}.
\item Level 3. Assessing the value of whole {\bf paradigms}, seen as collections of models.
\end{itemize}
In this hierarchy the Bayesian evidence at a given level acts both as the posterior's normalization at that level and as the likelihood function for the next level up. For example, at Level 1 we fix parameters (here collectively denoted by $\theta$) for a model $\cal M$ by maximising their posterior. Bayes' theorem reads: 
\be
P(\theta |D)=\frac{{\cal L}(\theta) \pi (\theta)}{P_{\cal M}(D)},
\ee
where ${\cal L}(\theta)=P(D|\theta)$ is the likelihood of the parameter $\theta$. The normalization of the posterior $P(\theta|D)$ is enforced by the denominator, which is also the evidence of $\cal M$ (i.e.~$P(D|{\cal M}$):
\be\label{evidM}
{\cal E}({\cal M})\equiv P(D|{\cal M})= P_{\cal M}(D)=\int {\cal L}(\theta) \pi (\theta)\, d\theta.
\ee
At the next level up (Level 2) we compare the relative merits of different models. Bayes' theorem now yields the posterior for model ${\cal M}$ within a paradigm $\mathcal{P}$:
\be\label{post-model}
P({\cal M}|D) = \frac{P(D|{\cal M}) \pi ({\cal M})}{P_{\cal P}(D)}.
\ee
As we see, at this level, the likelihood function (i.e.~the likelihood of the model ${\cal L}({\cal M})=P(D|{\cal M})$) is also the evidence of the model obtained from the preceding stage:
\be
{\cal L}({\cal M})= P(D|{\cal M})= {\cal E}({\cal M}).
\ee
This can be computed from level 1 according to Eq.~(\ref{evidM}).

Before proceeding to the next level up, we note that at this stage it is customary to evaluate the relative merits of two models, $\mathcal{M}_1$ and ${\cal M}_2$, within the same paradigm and given data $D$, by the ratio of their posteriors \footnote{The ratio can also be used if the models belong to different paradigms, but in this case the two posteriors have different normalization factors and normalization of the priors, so that Eq.~(\ref{ratio-model}) does not hold anymore.}. This can be written as:
\begin{equation}\label{ratio-model}
\frac{P(\mathcal{M}_1 | D)}{P(\mathcal{M}_2 | D)}=\frac{{\cal E}( \mathcal{M}_1)}{{\cal E}(\mathcal{M}_2)}\frac{\pi(\mathcal{M}_1)}{\pi(\mathcal{M}_2)}=B_{12}\frac{\pi(\mathcal{M}_1)}{\pi(\mathcal{M}_2)},
\end{equation}
where $B_{12}$ is the Bayes factor (the ratio of the evidences of models $\mathcal{M}_1$ and $\mathcal{M}_2$). When indifference between models is assumed (i.e.~when $\pi(\mathcal{M}_1)=\pi(\mathcal{M}_2)$) the ratio of the posterior probabilities is simply the Bayes factor. 

At Level 3 we evaluate the value of paradigm ${\cal P}$ by considering the whole set of models contained inside it. In fact, part of this exercise must be carried out if we want to normalise the posterior probability of its models, Eq.~(\ref{post-model}), although this is not needed for the evaluation of the ratio given by Eq.~(\ref{ratio-model}). The normalising denominator in Eq.~(\ref{post-model}) is given by:
\be
P_{\cal P}(D)=\int P(D|{\cal M})\pi ({\cal M})\, d{\cal M}=P(D|{\cal P}),
\ee
where the integral is made over all the models within paradigm $\mathcal{P}$, and the model priors are normalised to 1, i.e.~$\int \pi(\mathcal{M})d\mathcal{M}=1$. This is also the evidence of the paradigm, and it can be written as:
\be\label{EvidenceParadigm}
{\cal E}({\cal P})\equiv P(D|{\cal P})=\int {\cal E}({\cal M})\pi ({\cal M})\, d{\cal M}= P_{\cal P}(D).
\ee
Once again, the evidence of the paradigm acts as a normalization factor for the posterior probabilities of its models, but also as a likelihood function for the paradigm, should we want to compute its posterior probability. Indeed applying Bayes' theorem at the Level 3 gives us:
\be
P({\cal P}|D)=\frac{P(D|{\cal P}) \pi ({\cal P})}{P(D)}= \frac{{\cal E}({\cal P}) \pi ({\cal P})}{P(D)}.
\ee
We could speculate on the meaning of the normalization $P(D)$ at this stage but this is not necessary. As with models, we may assess the relative merits of two paradigms by considering the ratio of their posterior probabilities (from which $P(D)$ drops out):
\begin{equation}\label{newBfact}
\frac{P(\mathcal{P}_1 | D)}{P(\mathcal{P}_2 | D)}=\frac{{\cal E}( \mathcal{P}_1)}{{\cal E}(\mathcal{P}_2)}\frac{\pi(\mathcal{P}_1)}{\pi(\mathcal{P}_2)}.
\end{equation}
This would reduce to the Bayes factor between the paradigms were we to accept indifferent priors. However, as we shall now show, this is most inappropriate if we insist on predictivity being as relevant as fitting data. Bayesian evidence does penalise models for having too many parameters, but at the level of paradigm evaluation the penalty it imposes on lack of predictivity is not as harsh as the penalty for failing to fit data.

We summarise the results of this section in Table \ref{BayesLevels}.

\begin{table}[h!]
\centering
\begin{tabular}{| c || c || c |}
\hline
Levels
 & Likelihood & Posterior norm\\ \hline
Parameters & $P(D| \theta)$ & ${\color{blue!50!black}\mathcal{E}(\mathcal{M})=P(D|\mathcal{M})}$ \\ \hline
Models & ${\color{blue!50!black}\mathcal{E}(\mathcal{M})=P(D| \mathcal{M})}$ & ${\color{green!50!black}\mathcal{E}(\mathcal{P})=P(D| \mathcal{P})}$ \\ \hline
Paradigms & ${\color{green!50!black}\mathcal{E}(\mathcal{P})=P(D| \mathcal{P})}$ & $P(D)$ \\ \hline
\end{tabular}
\caption{\label{BayesLevels} Summary of likelihoods and posterior normalizations for three levels: parameters, models and paradigms. Colours show how the evidence is recycled into the next level up.}
\end{table}


\section{An illustrative extreme situation}\label{toy}
Let us consider a simple ``toy'' situation. Although the set up will seem a problem of model selection, we intend it as an illustration of the issues at stake in paradigm comparison. In fact, the concerns raised here are seldom relevant at Level 1 and 2 of the application of Bayes' theorem. 

\subsection{Paradigm testing vs. playing the lottery}\label{toy-example}
Consider two models dependent only on one parameter $\theta$: model $\mathcal{M}_1$ predicting its exact value, $\theta=\theta_p$, and model $\mathcal{M}_2$ leaving it entirely indeterminate. Let us assume that previous data or theoretical information require that $\theta_1<\theta<\theta_2$, so that $\mathcal{M}_2$ posits a uniform probability in this range. Within the Bayesian framework the two models are defined by priors upon $\theta$, with $\mathcal{M}_1$ associated with:
\be
\pi_1(\theta)=\delta(\theta-\theta_p),
\ee
and $\mathcal{M}_2$ with:
\be
\pi_2(\theta) = \left\{ \begin{array}{ll}
1/\Delta\theta & {\rm if} \quad \theta_1<\theta<\theta_2\\
 0 & {\rm otherwise} 
\end{array}
\right. , 
\ee
where $\Delta\theta=\theta_2-\theta_1$.

Let us assume for clarity that the data $\mathtt{d}$ results from signal $\theta$ and additive Gaussian noise $n$:
\be
\mathtt{d}=\theta+n,
\ee
and that $\sigma^2_N$ is the noise variance, but note that in general the data $\mathtt{d}$ need not be linear in the model parameters $\theta$ (e.g.~the inflationary models considered in \ref{PredInflationModels}). Then, the likelihood of parameter $\theta$ is:
\be\label{likelihood}
{\cal L}(\theta)=P(\mathtt{d}|\theta)=\frac{e^{-\frac{(\mathtt{d}-\theta)^2}{2\sigma_N^2}}} {\sqrt{2\pi}\sigma_N},
\ee
which peaks on $\theta=\mathtt{d}$ with variance $\sigma^2_N$. Let us further assume that $\sigma_N\ll \Delta\theta$, that is, the data (given the experimental sensitivity) is discriminative between models (if $\sigma_N\gg \Delta\theta$ models $\mathcal{M}_1$ and $\mathcal{M}_2$ will be indistinguishable). 
The evidence for the two toy models, according to Eq.~(\ref{evidM}), is:
\bea
{\cal E}_1 = {\cal E}(\mathcal{M}_1)& =& \frac{e^{-\frac{(\mathtt{d}-\theta_p)^2}{2\sigma_N^2}}} {\sqrt{2\pi}\sigma_N},\label{evid1}\\
{\cal E}_2 = {\cal E}(\mathcal{M}_2)&\approx &\frac{1}{\Delta\theta},\label{evid2}
\eea
where the last approximation is valid as long as $\mathtt{d}$ is comfortably in the range $\theta_1< \mathtt{d} <\theta_2$ (an approximation we shall drop later, but which is good enough here). The relative merits of these two models (posing here as ``toy paradigms'') are therefore assessed by the Bayes factor:
\be
\frac{P(\mathcal{M}_1 | D)}{P(\mathcal{M}_2 | D)}=B_{12}=\frac{e^{-\frac{(\mathtt{d}-\theta_p)^2}{2\sigma_N^2}}\Delta\theta} {\sqrt{2\pi}\sigma_N},
\ee
where we have assumed indifference, $\pi ({\cal M}_1)=\pi ({\cal M}_2)$, an assumption not to be confused with priors $\pi_i(\theta)$ defining the models themselves.

Of course, how well the models fare depends on the vicinity of $\mathtt{d}$ to $\theta_p$. However, the concept of evidence gives a softer penalization to lack of predictivity than to wrong predictions - in fact, it pays off to be un-predictive. If model $\mathcal{M}_1$ gets its prediction wrong it is penalised {\it exponentially}. In contrast, model $\mathcal{M}_2$ is penalised for not making a prediction merely as a {\it power-law}. We emphasise that this difference in penalisations is not related to the fact that one model has a Gaussian prior while the other has a flat prior. As we will see in Section \ref{IntermediateModel}, the same behaviour is observed for an intermediate model, and thus we conclude that this penalisation behaviour is the same regardless the particular form of the priors. 

These results are sensible from the point of view of probability theory: ``if you want to have a high probability of winning, then hedge your bets''. However, testing a paradigm is not about playing the lottery and winning, but instead about winning {\it given that you have bore the full brunt of potential loss}, by taking full chances of not winning {\it a priori}. In other words, we insist that testability of a paradigm should be considered as important as its capacity in fitting the data when assessing its success as a scientific theory. In particular, we want to propose an analysis procedure that can quantify how much of a paradigm's success in fitting the data is due to its capacity to adapt to any outcome of measurements and how much is instead due to it actually providing a correct description of the physical framework.  This is not well incorporated into the Bayesian evidence because the framework is designed for other ends, those of model selection rather than paradigm evaluation. Usually most models within a paradigm are equally predictive, so the issue need not arise at Level 2 (although this should be checked, e.g.~for inflation). 
\subsection{Identifying predictivity}\label{predictive}

In this paper we argue that within the framework of Bayesian evidence the issue of predictivity should be addressed by asking how well the theory would have fared {\it had the data been different}. It should not be enough for a successful  theory to exhibit a higher evidence than its competitors. The evidence should also be {\it exceptional with respect to what it would have been, had the data been different}. To quantify predictivity we propose to compute the probability distribution of the evidence given sampled randomly distributed data with given experimental noise. We illustrate this procedure with the two toy models introduced above. 

Let us assume that the sampled data $\mathtt{d}$ is within a ``reasonable'' range $\mathtt{d}_{1}<\mathtt{d}<\mathtt{d}_{2}$, and that we have a uniform distribution for $\mathtt{d}$ in this range, i.e.~$P(\mathtt{d})=1/\Delta \mathtt{d}$. We want to evaluate the induced probability distribution of the evidence ${\cal E}_i$ of the two models, given such randomly distributed hypothetical data (do not confuse these probabilities with those described in Section~\ref{3levels}, arising from the application of Bayes' theorem). 
For a fixed model the evidence is only a function of the data, ${\cal E}={\cal E}(\mathtt{d})$, called the {\it prior predictive distribution}. This will vary in a given range ${\cal E}_\text{min}<{\cal E}<{\cal E}_\text{max}$ as the data varies in $\mathtt{d}_{1}<\mathtt{d}<\mathtt{d}{_2}$. The induced probability distribution of ${\cal E}$ is given by:
\be\label{ProbE1}
P({\cal E})=\frac{1}{\Delta d }{\left|\frac{1}{{\cal E}'(\mathtt{d})}\right|},
\ee
when ${\cal E}_\text{min}<{\cal E}<{\cal E}_\text{max}$, and by zero otherwise. Here, the prime corresponds to a derivative with respect to $\mathtt{d}$. Notice that Eq.~(\ref{ProbE1}) is true if $\mathcal{E}(\mathtt{d})$ is a bijective function. However, most of the time $\mathcal{E}(\mathtt{d})$ will be a double-valued function, in which case Eq.~(\ref{ProbE1}) can be easily generalised as shown in Appendix \ref{App:ProbDouble}. 

\begin{figure}[h!]
\begin{center}
\scalebox{0.4}{\includegraphics{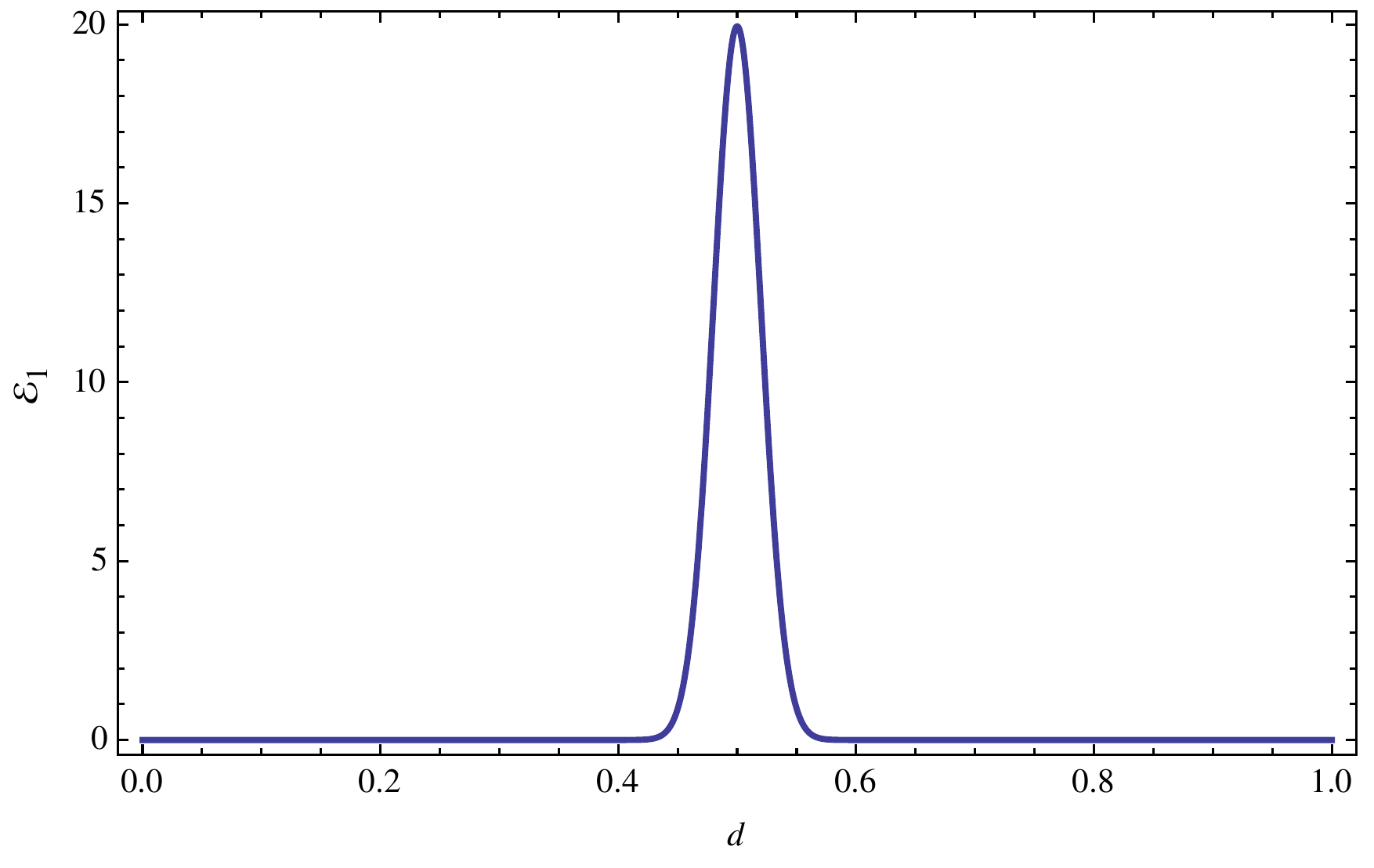}}
\caption{\label{fig:E2} The evidence as a function of the data for model $\mathcal{M}_1$, assuming
a prior with $\theta_p=0.5$, and a noise variance $\sigma_N=0.02$. As previously mentioned, the evidence will be high if the data is near the theoretical value $\theta_p$, but it will decay exponentially if the data is further away.}
\end{center}
\end{figure}

\begin{figure}[h!]
\begin{center}
\scalebox{0.4}{\includegraphics{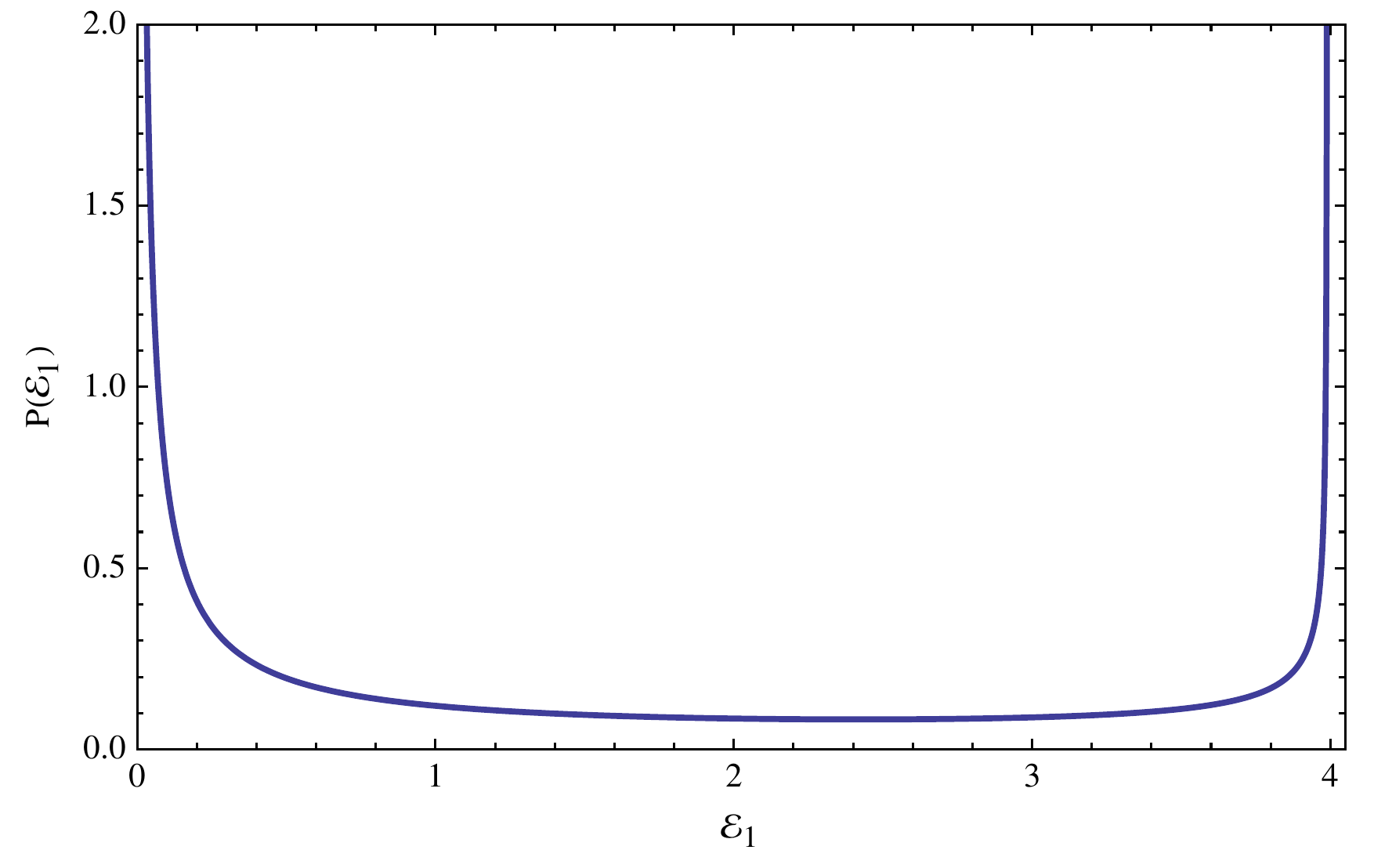}}
\scalebox{0.4}{\includegraphics{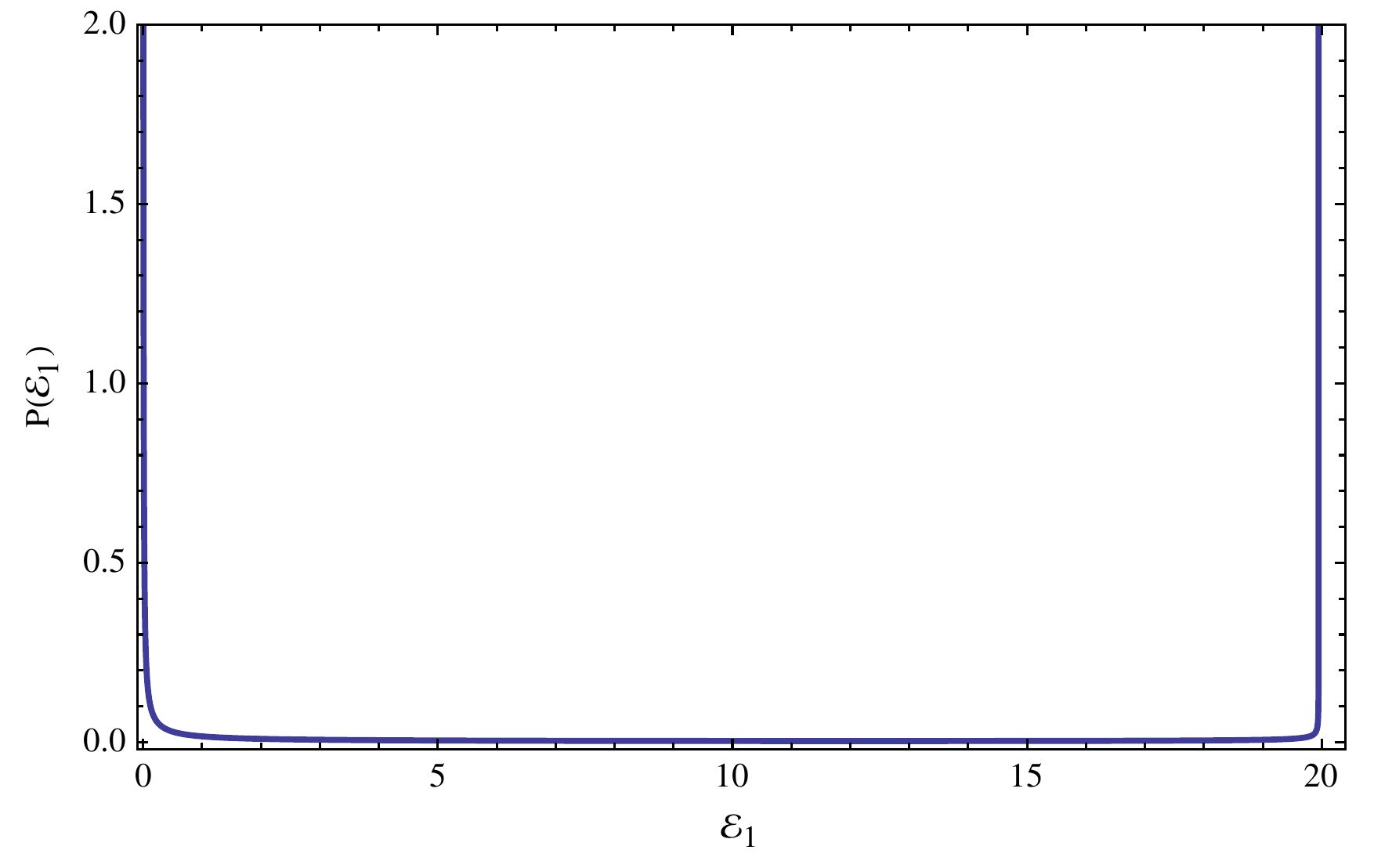}}
\caption{\label{fig:PE2}The distribution $P({\cal E})$ for model $\mathcal{M}_1$, assuming an interval $0<\mathtt{d}<1$ for the sampled data, a prior with $\theta_p=0.5$, and noise variance $\sigma_N=0.1$ (top) and $\sigma_N=0.02$ (bottom). As we see, as the noise decreases not only
the maximal evidence increases but also the percentage of the population living near the high peak decreases (in fact, proportionally
to $\sigma_N$).}
\end{center}
\end{figure}

As an illustration, we calculate explicitly the probability distribution of the evidence for the toys models described in Section~\ref{toy}. In order to obtain simple expressions, we will assume that the priors of both toy models are centred within the range of the sampled data. For model $\mathcal{M}_1$, we have plotted in Fig.~\ref{fig:E2} the evidence as a function of the data (with fixed variance). Assuming that $\theta_p$ is centred in the sampled data range, we will have for $\mathcal{M}_1$:
\bea
{\cal E}_\text{min}&=& \min ({\cal E}(\mathtt{d}_{1}),{\cal E}(\mathtt{d}_{2})), \\
{\cal E}_\text{max}&=&\frac{1}{\sqrt{2\pi}\sigma_N},
\eea
where now $\sigma_N$ is the noise of the sampled data (hypothetical experimental data). Thus, given Eq.~(\ref{evid1}), for ${\cal E}_\text{min}<{\cal E}<{\cal E}_\text{max}$ we find: 
\be
P({\cal E}_{1})=\frac{\sqrt{2}\sigma_N}{\Delta \mathtt{d}}\frac{1}{{\cal E}\sqrt{-\ln({\cal E}/{\cal E}_\text{max}})},
\ee
with $P({\cal E}_{1})=0$ outside this interval. The probability distribution is shown in Fig.~\ref{fig:PE2} for the particular case of $\mathtt{d}_{1}=0$ and $\mathtt{d}_{2}=1$. We can see that $P({\cal E}_{1})$ has two peaks, at the edges of the interval. The smaller the $\sigma_N$ the higher the ${\cal E}_\text{max}$, but also the smaller the percentage of the population living around the high evidence peak. As $\sigma_N\rightarrow 0$ almost all of the population lives at the low evidence peak, the hallmark of an ideal predictive theory.

\begin{figure}[h!]
\begin{center}
\scalebox{0.4}{\includegraphics{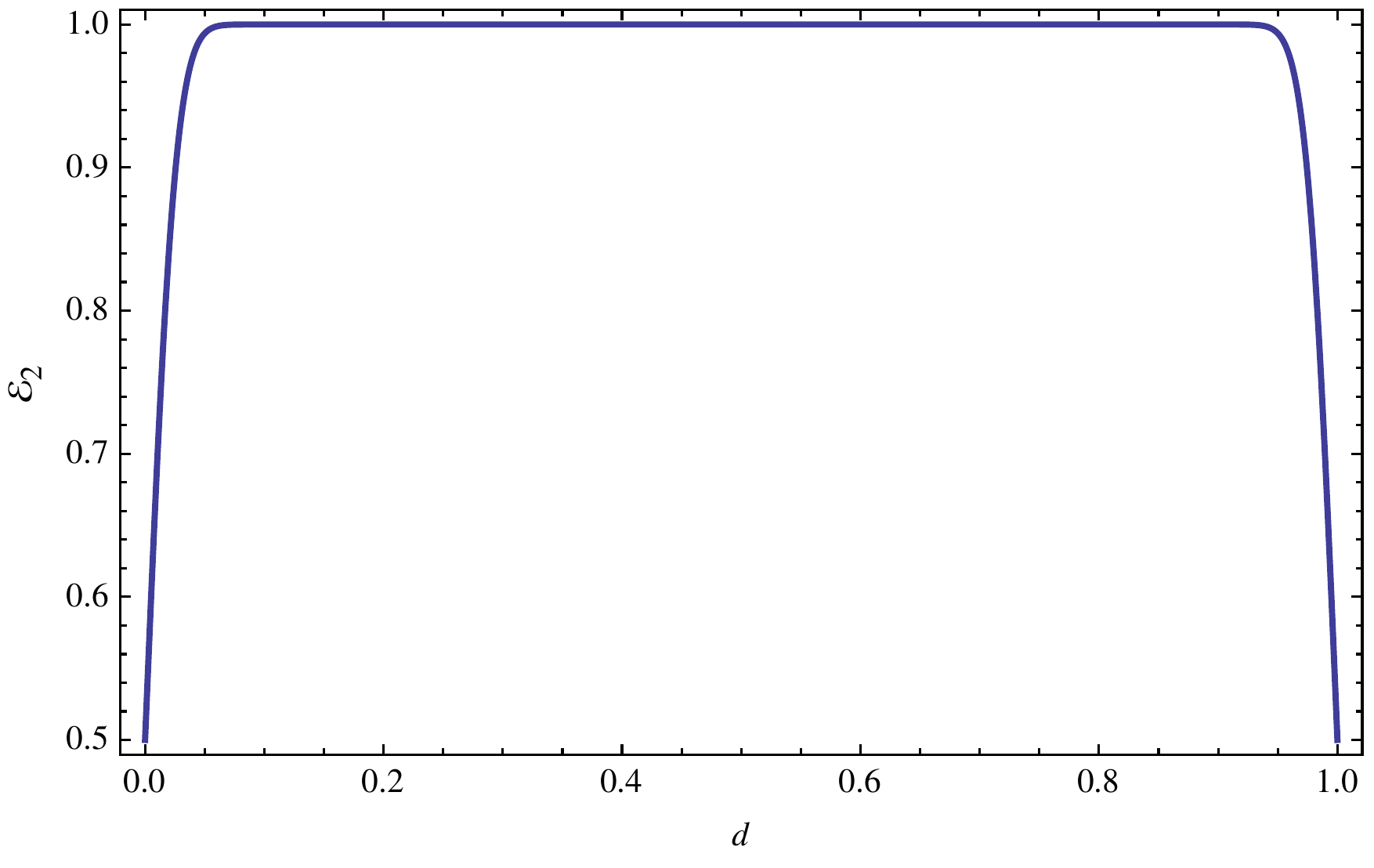}}
\caption{\label{fig:E1} The evidence as a function of the sampled data for model $\mathcal{M}_2$, assuming $0< \mathtt{d} <1$, and $\sigma_N=0.02$. As $\sigma_N$ decreases the evidence becomes more and more a constant within the interval, with a shaper step-function-like fall-off at the edges. }
\end{center}
\end{figure}

\begin{figure}[h!]
\begin{center}
\scalebox{0.4}{\includegraphics{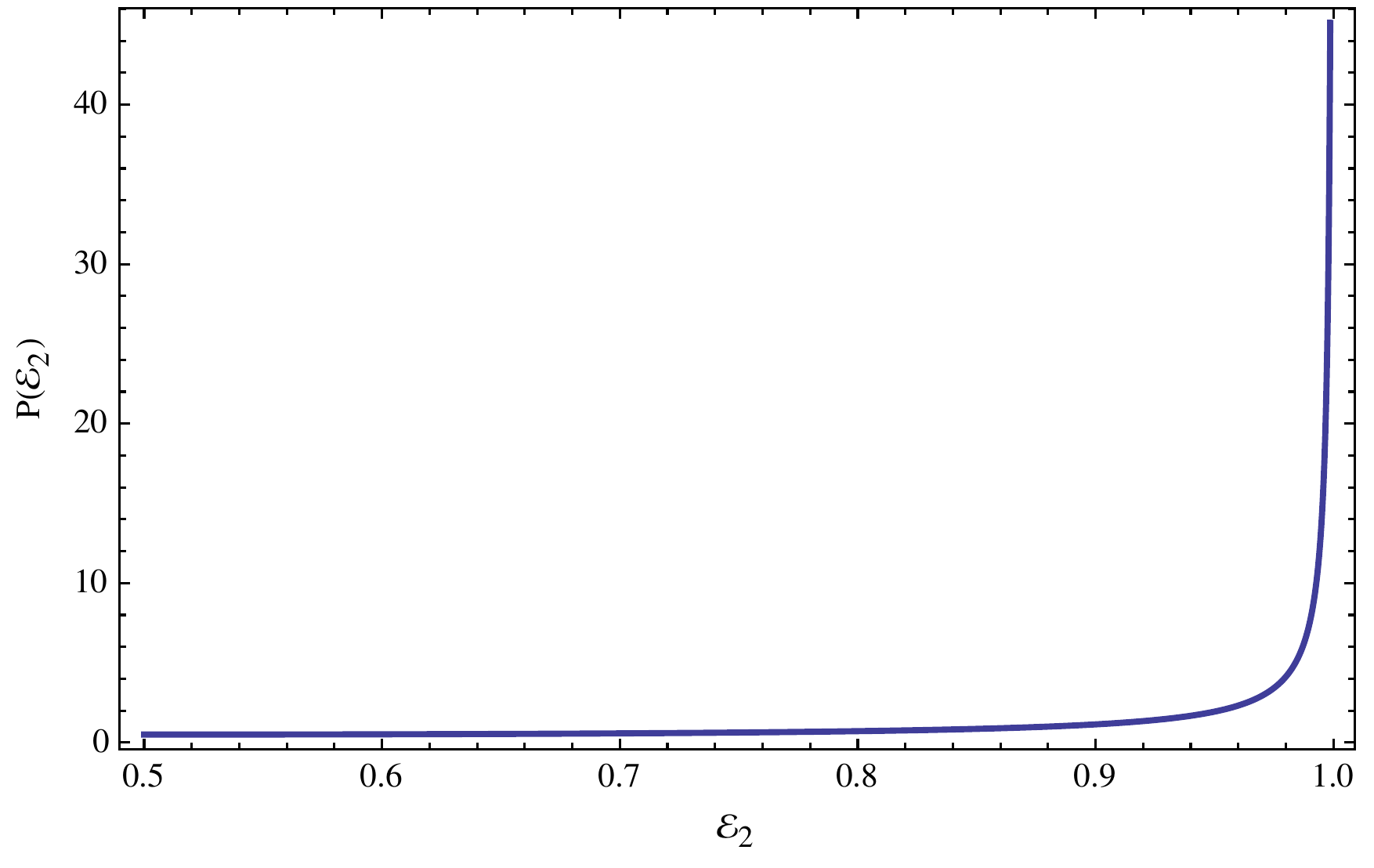}}
\scalebox{0.4}{\includegraphics{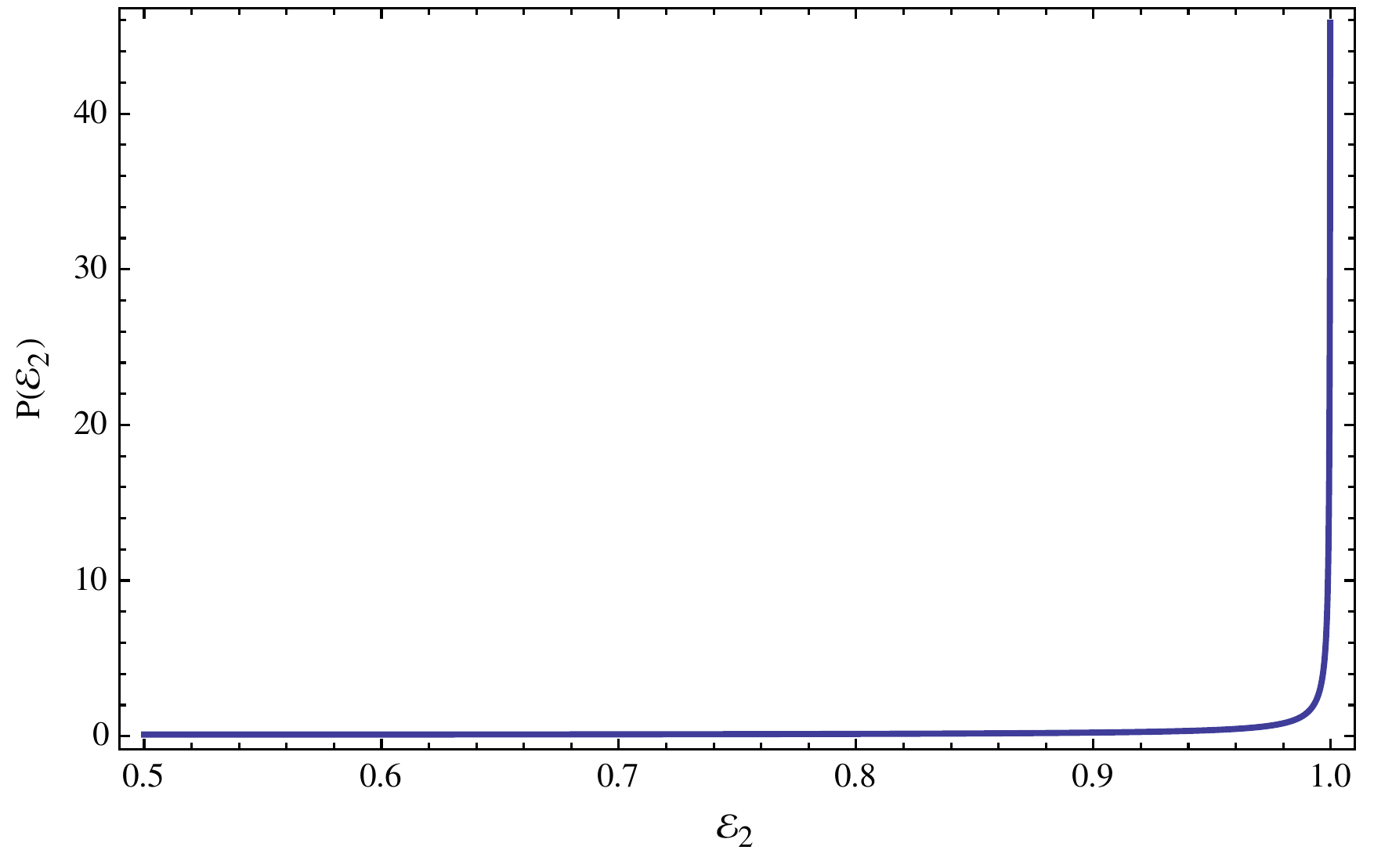}}
\caption{\label{fig:PE1}The distribution $P({\cal E})$ for model $\mathcal{M}_2$, assuming the same sampled data range as the prior $0< \mathtt{d} <1$, and $\sigma_N=0.1$ (top) and $\sigma_N=0.02$ (bottom). As we see in this extreme case only a high evidence peak exists, which becomes sharper and sharper as $\sigma_N\rightarrow 0$. }
\end{center}
\end{figure}

For model $\mathcal{M}_2$ we replace Eq.~(\ref{evid2}) by the exact expression:
\be\label{E2}
{\cal E}_2=\frac{
{\rm erf}\left(\frac{\mathtt{d}-\theta_1} {\sqrt{2}\sigma_N}\right)
-{\rm erf}\left(\frac{\mathtt{d}-\theta_2} {\sqrt{2}\sigma_N}\right)
}{2\Delta\theta},
\ee
which includes the fact that when $\mathtt{d}$ is close to the edges of the admissible range the evidence does vary, as illustrated in Fig.~\ref{fig:E1}. Assuming that the range of the prior $\Delta \theta$ is centred in the range $\Delta \mathtt{d}$, the evidence is such that:
\bea
{\cal E}_\text{min}&=& \min ({\cal E}(\mathtt{d}_1),{\cal E}(\mathtt{d}_2)), \\
{\cal E}_\text{max}&=&\frac{1}{\Delta\theta},
\eea
and so the probability distribution for ${\cal E}_\text{min}<{\cal E}<{\cal E}_\text{max}$ is given by:
\be
P({\cal E}_2)=\frac{2\sqrt{2\pi}\sigma_N\Delta \theta}
{
{\Delta \mathtt{d}\left|e^{-\frac{(\mathtt{d}-\theta_1)^2}{2\sigma_N^2}}-e^{-\frac{(\mathtt{d}-\theta_2)^2}{2\sigma_N^2}}\right|}
},
\ee
with $P(\mathcal{E}_2)=0$ outside this interval. This expression is to be understood as a parametric expression for $P({\cal E}_2)$ (with $\mathtt{d}$ the parameter) when taken in combination with Eq.~(\ref{E2}). Fig.~\ref{fig:PE1} shows the probability distribution for the particular case of $\mathtt{d}_1=\theta_1=0$ and $\mathtt{d}_2=\theta_2=1$, i.e.~$\Delta\theta=\Delta \mathtt{d}=1$. We can see that the distribution has a single peak at its maximal value $1/\Delta \theta$, which becomes sharper and sharper as $\sigma_N\rightarrow 0$. This is the hallmark of a totally unpredictive theory: the evidence does not depend much on the data; most of the population clusters at the high evidence peak. 

\subsection{Intermediate model}\label{IntermediateModel}
In this subsection 
 we set up a class of models with a parameter dialling between two extreme cases. If we add a theoretical noise $\sigma_T$ to model $\mathcal{M}_1$, we get what we call ${\cal M}_3$ described by the following prior:
\begin{equation}
\pi_3(\theta)=\frac{e^{-\frac{(\theta-\theta_p)^2}{2\sigma_T^2}}} {\sqrt{2\pi}\sigma_T},
\end{equation}
which gives the following evidence, when the likelihood is given by Eq.~(\ref{likelihood}):
\be
{\cal E}_3=\frac{e^{-\frac{(\mathtt{d}-\theta_p)^2}{2\sigma_{tot}^2}}} {\sqrt{2\pi}\sigma_{tot}},\label{evid3}
\ee
where $\sigma_{tot}^2=\sigma_N^2+\sigma_T^2$. From Eq.~(\ref{evid3}) we observe the same 
 penalisation previously mentioned. There is an exponential penalisation for not fitting well the data (i.e.~when $\theta_p$ is far from $\mathtt{d}$). On the contrary, being unpredictive is penalised at most as a power-law (for example when the theory fits the data very well, so that $(\mathtt{d}-\theta_p)< \sigma_{tot}$). We notice that these two penalisations are not simply due to a gaussian likelihood and a normalised prior, in fact, these two effects are hard to disentangle, and in some cases unpredictivity can even be rewarded, when the fit to the data is not perfect ($(\mathtt{d}-\theta_p)> \sigma_{N}$). Given the previous results for models $\mathcal{M}_1$ and $\mathcal{M}_2$, we conclude that the fact that there is a more severe penalisation for not fitting data than for being unpredictive in the standard Bayesian framework is quite general.

As for the previous models, we can find the probability distribution of the evidence for this model. First, we observe that in the range of the sampled data, the minimum and maximum evidences will be:
\bea
{\cal E}_\text{min}&=& \min ({\cal E}(\mathtt{d}_1),{\cal E}(\mathtt{d}_2)), \\
{\cal E}_\text{max}&=&\frac{1}{\sqrt{2\pi}\sigma_{tot}},
\eea
with a probability distribution of evidence for ${\cal E}_\text{min}<{\cal E}<{\cal E}_\text{max}$ given by: 
\be
P({\cal E})=\frac{\sqrt{2}\sigma_{tot}}{\Delta \mathtt{d}}\frac{1}{{\cal E}\sqrt{-\ln({\cal E}/{\cal E}_\text{max}})},
\ee
and $P({\cal E})=0$ outside this interval. Notice that these expressions are the same as for model $\mathcal{M}_1$, but now the noise present includes a theoretical noise $\sigma_T$. Thus, the behaviour of the probability distribution is the same, i.e.~has two peaks at the edges of the interval. The more predictive the theory (i.e.~smaller $\sigma_T$), the smaller the fraction of the population clustering around the high evidence peak. 

Notice that the hypothetical experimental noise $\sigma_{N}$ plays a similar role in determining the probability distribution. In fact the percentage of population around the high-evidence peak is of order $\sigma_{tot}/\Delta \mathtt{d}$, as we will rigorously show in the next section. So, if experiments provide bad-quality data ($\sigma_{N}\gg \sigma_{T}$) all theories will be similarly (un)predictive, even if the theoretical uncertainty is small, as the data does not allow falsifying models. Only with low experimental noise ($\sigma_{N}\ll \sigma_{T}$) predictivity will be determined by the theoretical uncertainty. As a consequence, the smaller the theoretical uncertainty, the better the data needs to be to pay justice to the model's predictive power.

In closing we note that we are developing a formalism for a concrete situation: paradigm testing given a continuous parameter and data variable, as well as non-vanishing noise. Several things might become ill-defined in other situations. Take for example testing whether a coin is fair or not with $N$ throws, which might seem like a counterpart of toy models 1 and 2, respectively. The evidence is exactly constant for the unfair coin model (counterpart of model 2), which renders the formalism degenerate. Noise is essential in removing pathologies from our framework: even though the situation described does 
have a continuous approximation, due to the central limit theorem in the large $N$ limit, in this approximation the likelihood is a Gaussian with noise given by $1/N$ and regrettably the noise vanishes at the same level of approximation where we can go to the continuous data limit. This is reflected in the fact that there are no edge effects in the discrete equivalent of model 2, whatever the value of $N$. These are essential in removing pathologies. We note that the motivation for our formalism is quite different from the problem just mentioned, so it is not surprising that we do not cover it. We are focusing on one specific problem: testing inflation and alternative paradigms using noisy CMB data.

\section{Predictivity defined}\label{pred-measure}
As the previous section shows, we can measure the predictivity of a model (or paradigm) by examining the distribution of the Bayesian evidence assuming uniformly distributed data. We want a predictive theory to have high evidence very rarely: we want high evidence to be
exceptional. 
One might think of employing a ``chi-squared measure'' for this feature, but the double-peaked nature of $P(\cal E)$ advises us against it. 
We propose instead using the 
percentage of the population living around the high peak as a measure of unpredictivity; the reciprocal population is then a measure of 
predictivity, $P_r$.

\subsection{A formula for the predictivity}
In general, we define the predictivity as:
\begin{equation}\label{Pred}
P_r=1-\int_{\bar{\cal E}}^{{\cal E}_\text{max}} P({\cal E}) d {\cal E},
\end{equation}
where the integral represents the population living between $\bar{\cal E}$ and ${\cal E}_\text{max}$. The predictivity varies between zero (perfectly unpredictive) and one (perfectly predictive), with the limiting cases found in the situations mentioned at the end of last section. The specific value of $\bar{\cal E}$ is arbitrary, and different values represent different measures of predictivity. In particular, for the choice $\bar{\cal E}={\cal E}_\text{max}/e$, this can be computed for model $\mathcal{M}_3$ as:
\be\label{Pr1}
P_{r3}=1-\frac{2\sqrt{2} \sigma_{tot}}{\Delta \mathtt{d}}.
\ee
Note that the last equation is a non-trivial result, and its derivation assumes that $\sigma_N,\sigma_T\ll \Delta \mathtt{d}$, 
and that $\theta_p$ is not close to the edges of the sampled interval. For model ${\cal M}_1$ the predictivity is given by:
\be\label{Pr1}
P_{r1}=1-\frac{2\sqrt{2} \sigma_{N}}{\Delta \mathtt{d}},
\ee
where we have assumed $\sigma_N\ll \Delta \mathtt{d}$, and that $\theta_p$ is not close to the edges of the sampled interval.

Naively, one might have expected this model to have the maximum predictivity, as its prior was a delta function. However, predictivity is a quantity that can only be defined after comparing with experimental uncertainty on data. For model $\mathcal{M}_1$ the predictivity is Eq.~(\ref{Pr1}), which can be close to one or zero, depending on $\sigma_N$. This happens because the experimental noise renders the high evidence region wider than expected from the prior (only a point in this case), making the model less predictive. This is a general feature. As data is improved, or new data becomes available, the predictivity of a model is expected to grow.

For model $\mathcal{M}_2$ we get the following predictivity:
\be
P_{r2}=1-\frac{\Delta\theta-2n\sigma_{N}}{\Delta \mathtt{d}},
\ee
where $n$ is a factor such that ${\cal E}(\theta_1+n\sigma_N)={\cal E}(\theta_2-n\sigma_N)={\cal E}_\text{max}/e$. In this expression we have assumed that the range $\Delta\theta$ is centred around the range $\Delta \mathtt{d}$. Note that for more unpredictive models, for example for a model $\mathcal{M}_3$ with 
\be\label{constr}
\sigma_{tot}> \frac{\Delta \mathtt{d}}{2\sqrt{2}},
\ee
according to Eq.~(\ref{Pred}), the predictivity will always be zero. 

We note that alternative definitions for predictivity could be defined other than Eq.~(\ref{Pred}). For example, the {\it information gain} $\mathcal{I}$ (or {\it Kullback-Leibler divergence}) \cite{kullback1951} measures the difference between the prior predictive distribution, given a model $\mathcal{M}$, relative to an assumed prior on the data $P(\mathtt{d})$:
\be\label{info}
\mathcal{I} = \int \mathcal{E}(\mathcal{M}) \ln\left( \frac{\mathcal{E}(\mathcal{M}) }{P(\mathtt{d})} \right) d\mathtt{d},
\ee
where it is implicitly assumed that $\mathcal{E}(\mathcal{M})$ depends on the data $\mathtt{d}$. One might then define predictivity as a monotonic function of the information $\mathcal{I}$, e.g.~$P_{r} \equiv 1 - e^{-\mathcal{I}}$. This form is motivated by the observation that the information gain measures the logarithmic ratio of the prior to posterior volumes.

\subsection{Alternative interpretation}

Let us consider a 1-dimensional case with a bijective prior predictive distribution. As previously stated, then, $P({\cal E}) |d{\cal E}|= p(\mathtt{d})\, |d\mathtt{d}|$, in which case predictivity can be re-written as:
\begin{align}\label{PrRewritten1}
P_r&=1-\int_{\bar{\cal E}}^{{\cal E}_\text{max}} P({\cal E}) d {\cal E},\nonumber\\
&=1- \int^{\theta_\text{max}}_{\bar{\theta}} P(\mathtt{d}) d\mathtt{d},\nonumber\\
&=1- \int^{\infty}_{-\infty}P(\mathtt{d})H(\mathcal{E}(\mathtt{d})-\bar{\mathcal{E}}) d\mathtt{d},
\end{align}
where $\bar{\theta}$ is such that $\mathcal{E}(\bar{\theta})=\bar{\mathcal{E}}$. In the last step we have introduced the function $H(x)$, which corresponds to the Heaviside step function such that $H(x)=1$ if $x\geq 0$ and $H(x)=0$ if $x<0$. In addition, if the sampled data is assumed to be uniformly distributed in the range $\Delta \mathtt{d}$, then $P(\mathtt{d})$ can be written as:
\begin{equation}\label{FlatData}
P(\mathtt{d})=\frac{H(\mathtt{d}_2-\mathtt{d})-H(\mathtt{d}_1-\mathtt{d})}{\Delta \mathtt{d}}.
\end{equation}
By replacing this into Eq.~(\ref{PrRewritten1}) we get:
\begin{align}\label{Pred2}
P_r&=1- \int^{\infty}_{-\infty}\frac{[H(\mathtt{d}_2-\mathtt{d})-H(\mathtt{d}_{1}-\mathtt{d})]}{\Delta \mathtt{d}}H(\mathcal{E}(\mathtt{d})-\bar{\mathcal{E}}) d\mathtt{d},\nonumber\\
&=1-\frac{1}{\Delta \mathtt{d}}\int^{\mathtt{d}_2}_{\mathtt{d}_1}H(\mathcal{E}(\mathtt{d})-\bar{\mathcal{E}}) d\mathtt{d}.
\end{align}
This expression for the predictivity is much simpler to calculate than that of Eq.~(\ref{Pred}), as it is not necessary to find $P(\mathcal{E})$. It also has a very clear interpretation: the predictivity of a model is the complement of the ratio of the range of data $\mathtt{d}$ ({\it within} the sampled data) giving evidence between ${\cal E}_\text{max}$ and $\bar{\cal E}$ and the full range of the sampled data. For instance, when $\bar{\cal E}={\cal E}_\text{max}/e$, for model $\mathcal{M}_1$, it is clear that the range of the data $\mathtt{d}$ giving an evidence between ${\cal E}_\text{max}$ and ${\cal E}_\text{max}/e$ is $2\sqrt{2}\sigma_{N}$ (with the assumptions previously mentioned). The predictivity of the model is then that given by Eq.~(\ref{Pr1}). 

Equation (\ref{Pred2}) is general, and its derivation can be found in Appendix \ref{App:PredDouble} for a 1-dimensional case with a double-valued prior predictive distribution, and in Appendix \ref{App:PredNdim} for $N$ dimensions.



\section{Folding predictivity into bayesian framework}\label{new-factor}
We now return to Section~\ref{3levels} to re-assess the priors that should be inserted into Eq.~(\ref{ratio-model}) and Eq.~(\ref{newBfact}) in order to incorporate predictivity into the Bayes factor for models and paradigms. When using modified priors, we will be ultimately interested in the ratio of posterior probabilities given the data.
We want to use a prior that gives due value to predictivity. In general, any function of the predictivity could be used as prior. In this paper we propose to give at least the same importance to predictivity as to goodness of fit, i.e.~to penalise lack of predictivity at least as severely as the Bayesian evidence penalises models for failing to fit the data. 

Let us return to the toy models presented in Section \ref{toy} and note how model $\mathcal{M}_1$ is penalised exponentially for not-fitting the data. 
We thus argue that model ${\cal M}_2$ should be exponentially penalised for its lack of predictivity. One way to enforce this is to select the prior:
\be
\pi(\mathcal{M})=\frac{1}{N}\; e^{-(1-P_r)^2/P_r^2},\label{PriorPr}
\ee
where $N$ is a normalization factor such that $\sum_i \pi(\mathcal{M}_i)=1$. We can see that for a perfectly unpredictive model, ($P_r=0$) this prior is zero, while for a perfectly predictive model ($P_r=1$) it becomes $\frac{1}{N}$. As required, as $P_r\rightarrow 0$ the prior vanishes
exponentially.

We can easily compute the prior for model ${\cal M}_3$:
\be
\pi({\cal M}_3)=
\frac{1}{N} e^{-\frac{8\sigma_{tot}^2}{(\Delta \mathtt{d}-2\sqrt{2}\sigma_{tot})^2}}.
\ee
As we see, $\pi({\cal M}_3)\rightarrow 0$ exponentially as $\sigma_{tot}\rightarrow \Delta \mathtt{d} /(2\sqrt{2})$, in the same way as $\mathcal{E}_3\rightarrow 0$ when $|\mathtt{d}-\theta_p|\rightarrow \infty$. Therefore, with this prior we give the same penalisation for not being predictive as for not fitting the data. Analogously, model ${\cal M}_1$ receives the prior:
\be\label{PriorM1}
\pi({\cal M}_1)=\frac{1}{N}
e^{-\frac{8\sigma_{N}^2}{\Delta \mathtt{d}^2}}\rightarrow \frac{1}{N} \quad \mbox{if} \; \sigma_N\ll \Delta \mathtt{d}.
\ee
 For model ${\cal M}_2$ we get:
\be\label{PriorM2}
\pi({\cal M}_2)=\frac{1}{N}
e^{-\frac{(\Delta\theta-2n\sigma_{N})^2}{[(\Delta \mathtt{d}-\Delta\theta)+2n\sigma_N]^2}}\approx 
\frac{e^{-\frac{(\Delta\theta)^2}{(4n\sigma_N)^2}}}{N}\rightarrow 0,
\ee
where we have assumed that $\Delta\theta \gg \sigma_N$, and that $(\Delta \mathtt{d}-\Delta\theta)\sim 2n\sigma_N$. Then, the ratio of the posterior probabilities between models $\mathcal{M}_1$ and $\mathcal{M}_2$ is: 
\begin{equation}
\frac{P(\mathcal{M}_1|D)}{P(\mathcal{M}_2|D)}\approx B_{12}
e^{-\frac{(\Delta\theta)^2}{(4n\sigma_N)^2}}.
\end{equation}

As it was already noted in the previous sections (see also appendix \ref{App:ExNoise}), the prior we are proposing depends on the experimental errors. Even though this is unusual for priors, the fact that this prior is meant to assess how predictive a theory is makes the noise dependence a necessary feature: the level of predictivity of a theory is determined from what can be reasonably expected to be measurable. In the same way, predictivity will also depend on what the observables are, and if experimental advances allow access to a new set of observables, theories will have to be confronted with these as well. We stress though that predictivity does not depend on the data, as the level of experimental noise and possible set of observables can be known when an experiment is set up, before getting any data.

We also remark that in \cite{2011MNRAS.413.2895J,2014AIPC.1636..106J} it was also argued that for assessing the performance of a model not only the value for the evidence should be considered, but the distribution of the evidence for hypothetical data as well. It was shown that the Bayes factor is a noisy statistic, as it is strongly affected by the signal to noise ratio in the data and the assumed priors. Thus, it was concluded that even though the Bayes factor expresses
our knowledge of the relative odds on different models, we need to consider more than a single numerical value for making a final decision. 

Finally, we stress that our formalism adds a new, distinct level of subjectivity to the usual one, frequently recognised within the Bayesian framework. The usual problem concerns the ambiguity in the choice of measure (and concomitant priors) over the space of parameters associated with theoretical models. In setting up our predictivity prior we do require a model-independent measure on the space of the data itself, and this carries with it a separate ambiguity. The two problems formally mimic each other, yet it is important to stress that we are indeed adding a new layer of subjectivity, because the data measure is model-independent, and so independent of the priors placed upon the model parameters themselves.

\subsection{A toy numerical example}
Let us consider a numeric example drawn from the two toy models $\mathcal{M}_1$ and $\mathcal{M}_2$. Let the data be $\mathtt{d}=0.4$, with noise variance $\sigma_N=0.02$. If the prior of the models are such that $\theta_p=0.5$ and $\Delta \theta=1$, then the Bayes factor $B_{12}$ is:
\begin{equation}
\ln(B_{12})=-9.5,
\end{equation}
which corresponds to a strong evidence for model $\mathcal{M}_2$ over $\mathcal{M}_1$, according to the Jeffrey's scale shown in Table \ref{JeffScale}. To be unpredictive pays off, if we rely solely on the Bayes factor. 
\begin{table}[h]
\centering
\begin{tabular}{| c || c || c |}
\hline
$\ln(B_{12})$
 & Odds & Strength of evidence\\ \hline
$<1.0$ & $\lesssim 3:1$ & Inconclusive \\ \hline
$1.0 - 3.0$ & $ 3:1-20:1$ & weak evidence \\ \hline
$3.0 - 5.0$ & $ 20:1-150:1$ & moderate evidence \\ \hline
$>5.0$ & $\gtrsim 150:1$ & strong evidence \\ \hline
\end{tabular}
\caption{\label{JeffScale} Jeffreys' scale for translating values of Bayes factors into strengths of evidence, when comparing two models $\mathcal{M}_1$ versus $\mathcal{M}_2$.}
\end{table}

Let us now calculate the posterior probabilities of the models when including the prior given by Eq.~(\ref{PriorPr}). If the range of the sampled data for calculating the predictivity is $\Delta \mathtt{d}=1+10 \sigma_N$ (with the range of model $\mathcal{M}_2$ prior inside this range), then the predictivities become:
\begin{equation}
P_{r1}=0.953; \quad P_{r2}=0.054.
\end{equation}
As expected, model $\mathcal{M}_1$ is very predictive as $P_{r1}\sim 1$, while model $\mathcal{M}_2$ is much less predictive. The ratio of the posterior probabilities become:
\begin{equation}
\ln\left(\frac{P(\mathcal{M}_1|D)}{P(\mathcal{M}_2|D)}\right)=297.4,
\end{equation}
resulting in a strong evidence for model $\mathcal{M}_1$ over $\mathcal{M}_2$, according to Jeffrey's scale. We can see that by giving due importance to predictivity, the preference of a model can change dramatically. In this example, model $\mathcal{M}_2$ fits better the data than model $\mathcal{M}_1$, because it has a higher evidence. With only this factor in consideration, we would prefer model $\mathcal{M}_2$. However, this model is considerably less predictive, resulting in a preference over model $\mathcal{M}_2$ on the final posterior probabilities. 

It is important to mention that this result depends highly on the range of sampled data considered for calculating the predictivities. The larger the range, the better the predictivities, as seen in Eq.~(\ref{PriorM1}) and (\ref{PriorM2}). This is why it is important to choose a reasonable sampled data range. This ambiguity is analogous to that present in the standard Bayesian formalism where some reasonable range for flat priors on parameters must be placed and chosen on a case-by-case basis.

\subsection{Paradigm vs model comparison}
In this subsection we illustrate how the issue of predictivity may be strongly contextual and change radically between Level 2 (model comparison) and 3 (paradigm comparison). When comparing models, flat priors are usually employed, in line with expected comparable predictivities. However, the fact that paradigms are collections of models does not preclude the relevance of predictivity, even when the individual models are roughly equally predictive. 


In general we would like to impose the same philosophy regarding predictivity to models and paradigms. Thus, priors for paradigms and models should have the same form:
\be\label{PriorParadigm}
\pi(\mathcal{P})=\frac{1}{N}\; e^{-(1-P_r)^2/P_r^2},
\ee 
where $N$ is some normalization, and the predictivity $P_r$ is now calculated using the evidence of the paradigm given by Eq.~(\ref{EvidenceParadigm}). In principle, the form of the prior for paradigms, $\pi({\mathcal P})$, and for models, $\pi({\mathcal M})$, could be different. However, we shall not consider this possibility here. 

\begin{figure}[h]
\begin{center}
\scalebox{0.4}{\includegraphics{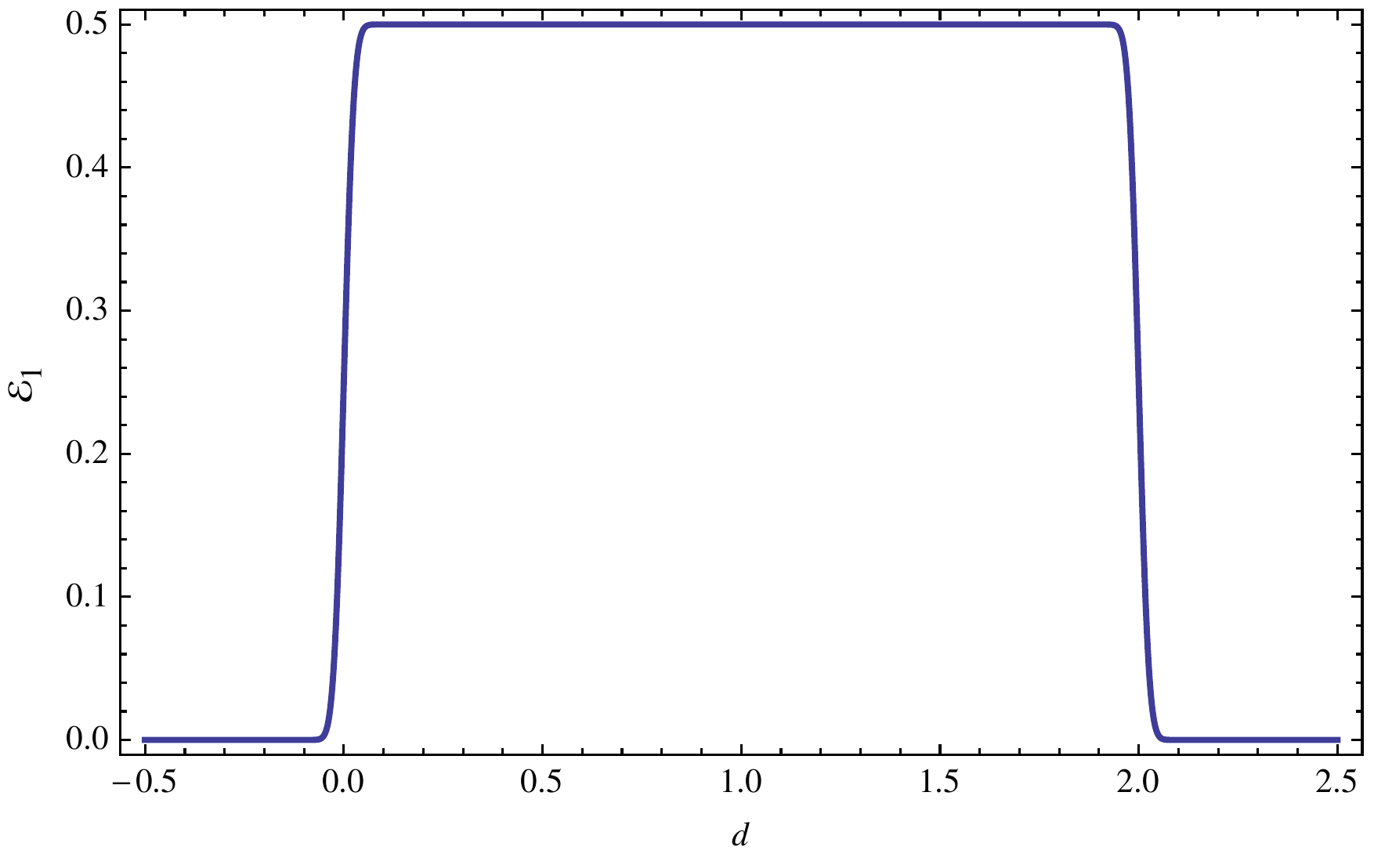}}
\caption{\label{fig:Paradigm1} The evidence as a function of the data for paradigm $\mathcal{P}_1$, assuming $\sigma_N=0.02$.}
\end{center}
\end{figure}
\begin{figure}[h]
\begin{center}
\scalebox{0.41}{\includegraphics{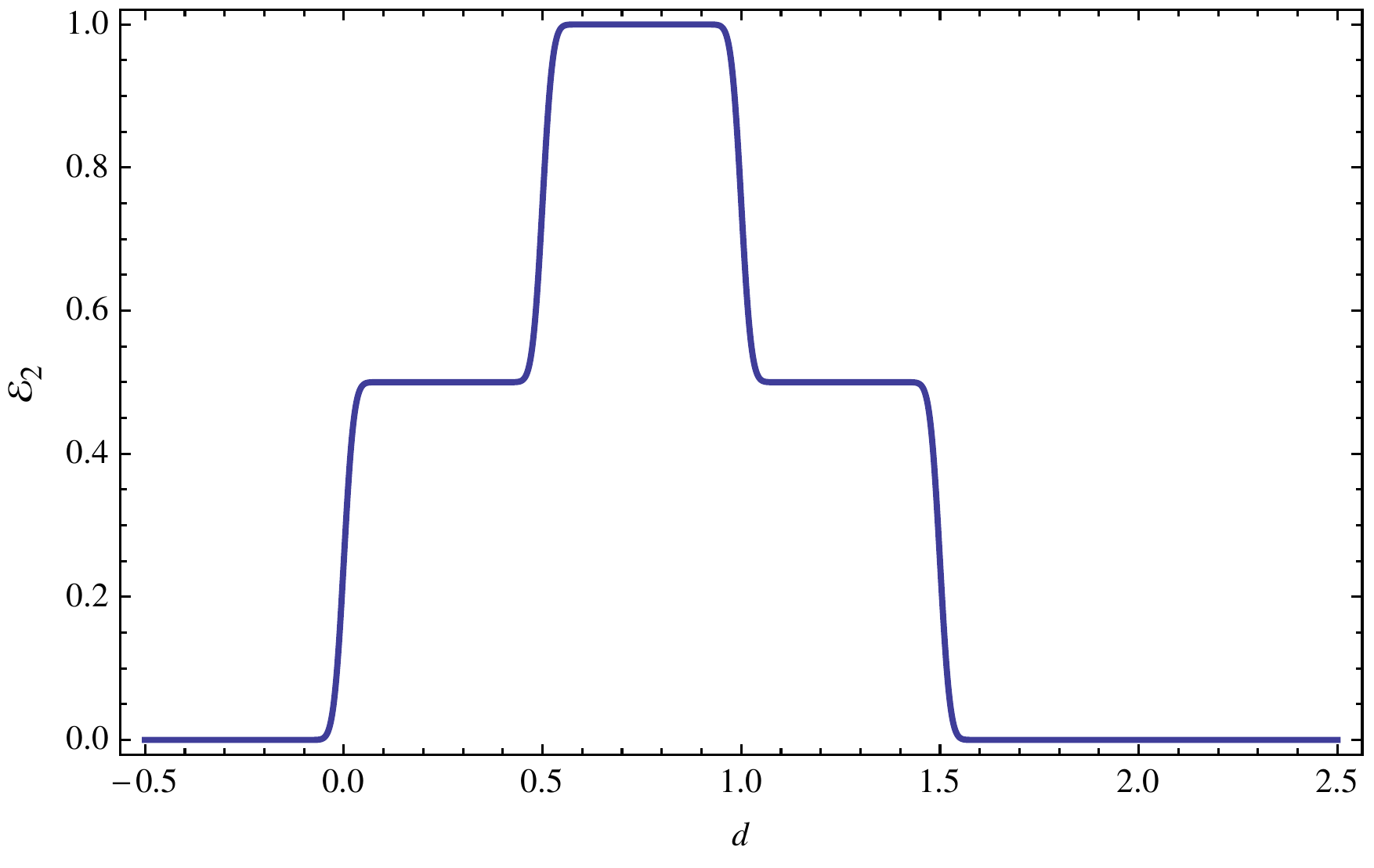}}
\caption{\label{fig:Paradigm2} The evidence as a function of the data for paradigm $\mathcal{P}_2$, assuming $\sigma_N=0.02$.}
\end{center}
\end{figure}

We now show with a numerical example how the issue of predictivity may be irrelevant at Level 2 (model selection) but not at Level 3 (paradigm evaluation). Consider two paradigms each made up of two models: $\mathcal{P}_1=\{\mathcal{M}_1,\mathcal{M}_2\}$ and $\mathcal{P}_2=\{\mathcal{M}_3,\mathcal{M}_4\}$, where the four models result from flat priors upon their free parameter $\theta$ given by: 
\begin{align}
&\pi_1(\theta) =\pi_3(\theta)= \left\{ \begin{array}{ll}
1/\Delta\theta & {\rm if} \quad 0<\theta<1\\
 0 & {\rm otherwise} 
\end{array}
\right. , \\
&\pi_2(\theta) = \left\{ \begin{array}{ll}
1/\Delta\theta & {\rm if} \quad 1<\theta<2\\
 0 & {\rm otherwise} 
\end{array}
\right. , \\
&\pi_4(\theta) = \left\{ \begin{array}{ll}
1/\Delta\theta & {\rm if} \quad 0.5<\theta<1.5\\
 0 & {\rm otherwise} 
\end{array}
\right. . 
\end{align}
The salient feature is that all four models have the same $\Delta\theta=1$, but in different regions of $\theta$. Therefore all the models have the same predictivity, and so the ratio of their posterior probabilities is simply the corresponding Bayes factor. Now, let us calculate the predictivity of each paradigm when the sampled data lies in $\mathtt{d}=[-0.5,2.5]$, with a fixed noise variance $\sigma_N=0.02$. In Figures \ref{fig:Paradigm1} and \ref{fig:Paradigm2} we plot the evidence as a function of the data for paradigms $\mathcal{P}_1$ and $\mathcal{P}_2$, respectively.
In this numerical example we obtain:
\begin{equation}
P_{r\mathcal{P}_1}=0.33; \quad P_{r\mathcal{P}_2}=0.51, 
\end{equation}
and using Eq.~(\ref{PriorParadigm}), this gives:
\begin{equation}
\frac{\pi(\mathcal{P}_1)}{\pi(\mathcal{P}_2)}=4\times 10^{-2}.
\end{equation}
Therefore, according to Eq.~(\ref{newBfact}), the ratio of the posterior probabilities of the paradigms changes by two orders of magnitude if we do take predictivity into account. And yet all the models considered, making up the paradigms, have the same predictivity, $P_r=0.62$. This example shows clear that even when the standard procedure of ignoring predictivity at Level 2 is correct, this is no reason to forego a more thorough  analysis when assessing paradigms, at Level 3. 


We finally remark that since model priors are always normalised to 1, the evidence of a paradigm is not determined by the number of models within it, and hence nor its predictivity. Instead, the predictivity of a paradigm will depend on the overall spread of predictions made by its models. For this reason, we can compare all kind of paradigms regardless the number of models within them.

\section{Predictivity of inflationary models}\label{PredInflationModels}

In this section we apply the concepts developed in the previous sections to inflationary models. We consider a large set of inflationary models, find their Bayes factors when compared to a reference model, evaluate their predictivities, and compute their ratios of posterior probabilities once the predictivity is incorporated into the priors. We stress that in defining the predictivity a choice of data ``$D$" is assumed. In general, for different choices of ``$D$'' (i.e~different sets of observables), there will be different distributions for the evidence, and thus different predictivities. Since we are focusing on inflationary models, we will assume the observables to be the tensor-to-scalar ratio amplitude of primordial perturbations $r$, and the spectral index for primordial scalar perturbations $n_\text{s}$. 

This is an important assumption as, in principle, we could have used a completely different set of parameters as data, such as the power spectrum multipole moments of the Cosmic Microwave background (CMB) temperature $C_{\ell}$. This ambiguity mimics that found when deciding on what parameters a prior should be assumed when calculating the evidence of a model in the Bayesian framework. 

In fact, when in Section~\ref{3levels} we examined chains of applications of Bayes' theorem, we neglected to point out that a level 0 (and several sub zero levels) could have been added to the chain, since quite often what is seen as a ``parameter'' at one level becomes ``data'' at another. This is a common situation in data reduction: no one ever compares the theory with raw data (presumably the time-ordered series in CMB temperature experiments). This is converted into a chain going from the time-ordered series, to a map (or $a_{\ell m}$), to the $C_\ell$ and to the cosmological parameters. We could have called ``data''  the input from any of these levels. Since we are working in the same Bayesian framework, we carry this ambiguity to the definition of predictivity. For simplicity, we will consider only two observables, $r$ and $n_\text{s}$, with all the other parameters marginalised\footnote{The amplitude of primordial scalar perturbations is not a discriminative observable, as all inflationary models considered here have the same dependence on this parameter, and the same prior. Thus, if we added this parameter as data, all the predictivities would change in the same way. However, by adding CMB non-Gaussian observables as data, we might see a difference in the relative predictivity of models, but we do not add them here as the computational effort would increase considerbaly.}.


\subsection{Inflationary models}

For simplicity, we consider only the 1D and 2D single-field slow-roll inflationary models described in \cite{Martin:2013nzq} (see also \cite{Martin:2014vha} for details). Thus, we consider a total of 85 models out of the 193 described in \cite{Martin:2013nzq}. Throughout this section we use the same choice of priors and terminology (including model initials) as in \cite{Martin:2013nzq}. Each model is defined by a particular inflationary potential and parameter priors. This way, two models can have the same inflationary potential and differ only in the priors choice for its parameters.

Specifically, the 85 models in our analysis all have a reheating parameter and at most one relevant potential parameter. For instance, one 1D model is Higgs inflation (HI). This model has the following potential:
\begin{equation}
V(\phi)=M^4\left( 1- e^{-\sqrt{2/3}\phi/M_{p}}\right)^2,
\end{equation}
where $\phi$ is the inflaton field, $M$ is some mass scale for the potential and $M_p$ is the Planck mass. We say that this model is 1D as we are interested, as observables, in the tensor-to-scalar ratio $r$ and the scalar spectral index $ n_\text{s}$, which do not depend on the mass scale $M$, but only on a reheating parameter\footnote{If we were to consider the primordial scalar perturbation amplitude as an observable, we would also have to include $M$ as a relevant free parameter in the model, as the two are directly related. However, since we are only interested in $r$ and $ n_\text{s}$, $M$ is marginalized in calculations.}. This reheating parameter will be defined as:
\begin{equation}
R\equiv R_{\text{rad}}\frac{\rho_{\text{end}}^{1/4}}{M_p}; \quad R_{\text{rad}}=\frac{a_{\text{end}}}{a_{\text{reh}}}\left(\frac{\rho_{\text{end}}}{\rho_{\text{reh}}}\right),
\end{equation}
where $\rho_{\text{reh}}$ and $\rho_{\text{end}}$ correspond to the energy density at the end of the reheating era and the end of inflation, respectively. Similarly, $a_{\text{reh}}$ and $a_{\text{end}}$ correspond to the scale factor at the end of the reheating era and the end of inflation, respectively. Notice that $R$ is dimensionless. 

As we stated, a model is also defined by the choice of prior. Following~\cite{Martin:2013nzq}, we choose the prior for the reheating parameter $R$ to be uniform in the logarithm of $R$:
\begin{equation}\label{RPrior}
\pi\left(\ln (R)\right)=U(-46,15).
\end{equation}
This prior covers all reheating histories satisfying that the mean equation of state $w_{\text{reh}}$ during reheating is $-1/3 < w_{\text{reh}}<1$, and
$\rho_{\text{nuc}}<\rho_{\text{reh}}<\rho_{\text{end}}$, where $\rho_{\text{nuc}}$ is the energy density at Big Bang nucleosynthesis (i.e.~it satisfies $\rho_{\text{nuc}}^{1/4}=10\text{MeV}$). Eq.~(\ref{RPrior}) will be the prior on $R$ for all models.

Two-dimensional models have an extra relevant parameter, related to the the inflationary potential, and a flat prior will be assumed on it. As an example, consider $R+R^{2p}$ inflation 3 (RPI3). This model has the following potential: 
\begin{equation}
V(\phi)=M^4e^{-2\sqrt{2/3}\phi/M_p} \rvert e^{\sqrt{2/3}\phi/M_p}-1 \lvert^{2p/(2p-1)},
\end{equation}
where $\phi$ is the inflationary field, $M$ is some mass scale, $M_p$ is the Planck mass, and $p$ some free index determining the shape of the potential. In this case, the observables $r$ and $ n_\text{s}$ depend on two parameters: the reheating parameter $R$, and the index $p$. Notice that the case with $p=1$ corresponds to HI. RPI3 is defined by the prior choice given by Eq.~(\ref{RPrior}), and a flat prior on $p$ in the range:
\begin{equation}
\pi(p)=U(0.8,1).
\end{equation}
For details on other models and their priors, we refer the reader to \cite{Martin:2013nzq} and \cite{Martin:2014vha}.

\subsection{Methodology}

We calculate the Bayesian evidence for all 1 and 2 parameter models in the ASPIC inflationary model library \cite{Martin:2014vha}, given the Planck 2015 data \cite{PlanckXVI}. As explained, we choose prior $\pi$ on $R$ according to Eq.~(\ref{RPrior}) and a flat prior on the potential parameter $\theta_{\rm inf}$, if relevant. We calculate the likelihood on a high-resolution grid in this parameter space by calculating the corresponding $(r, n_\text{s})$ pair, given the potential and reheating parameters, and substituting this into a marginalised Planck likelihood $\mathcal{L}[r, n_\text{s}]$. The evidence is then simply calculated as:

\begin{equation}
\mathcal{E} = \int dR d\theta_{\rm inf} \mathcal{L}[ r(R, \theta_{\rm inf}), n_\text{s}(R, \theta_{\rm inf}) ] \pi(R) \pi(\theta_{\rm inf}).
\end{equation}

The marginalised Planck likelihood $\mathcal{L}[r, n_\text{s}]$ is constructed by building a histogram in the $(r, n_\text{s})$-plane of the publicly-available Planck likelihood samples (the MCMC chain) \cite{Ade:2013kta}. This procedure implicitly marginalises over all other $\Lambda$CDM parameters, including the primordial scalar perturbation amplitude $A_s$. For simplicity and computational speed we make a normal approximation to the likelihood by fitting a Gaussian to this histogram. We find the best-fit parameters
\begin{equation}
\label{bestFitrns}
\bar{r}=-0.077; \quad \bar{n}_s=0.964,
\end{equation}
and covariance such that
\begin{equation}\label{CorrMatrix}
\sigma_r=0.075; \quad \sigma_{n_\text{s}}=0.007;\quad \rho_{r, n_\text{s}} = 0.196,
\end{equation}
where $\sigma_r$, $\sigma_{n_\text{s}}$ is the Gaussian width on $r$ and $n_\text{s}$, respectively, and $\rho_{r, n_\text{s}}$ their correlation coefficient. Our results on the evidence of models compare closely to \cite{Martin:2013nzq} (see Appendix \ref{App:PredInflModels}), as expected given the Planck 2015 likelihood in $r, n_\text{s}$ is very similar to the 2013 likelihood. We use Higgs Inflation (HI) as the reference model for ease of comparison.

In addition, we calculate the predictivities for each model $\mathcal{M}_i$ as follows. The prior predictive distribution $\mathcal{E}(r, n_\text{s}| \mathcal{M}_i)$ is computed for each $(r, n_\text{s})$ pair on a grid defined over $r\in[0,0.45]$ and $n_\text{s}\in[0.92,1.01]$, corresponding to $\sim 7\sigma$ from the best-fit Planck values. This amounts to recomputing the Bayesian evidence where now the {\it data} are a translated version of the Planck likelihood: each $(r, n_\text{s})$ pair defines hypothetical data which are {\it a priori} plausible and replace the best-fit values in Eq.~(\ref{bestFitrns}). We fix the covariance matrix of the sampled data to be given by Eq.~(\ref{CorrMatrix}). This produces a density plot of the prior predictive distribution for each model (e.g.~Fig.~\ref{fig:PredHI} for HI). The predictivity is then given by Eq.~(\ref{Pred2}): the ratio of the area of the data plane such that $\mathcal{E} > \mathcal{E}_\text{max}/8e$ to the full area. Models with diffuse prior predictive distributions result in a posterior evidence which is weakly-dependent on the data and hence are assigned low predictivities, and vice versa. 

Finally, we incorporate predictivity into the Bayesian paradigm by inserting the predictivities into Eq.~(\ref{PriorPr}) to define a model prior probability which encapsulates the predictivity of each model. This allows us to calculate the posterior probabilities for each model using Eq.~(\ref{ratio-model}) (with HI as the reference model). 

To close this subsection, we briefly justify some of our technical assumptions, referring the reader to the Appendices for more detail. We use a Gaussian approximation of the Planck likelihood because the computation of the prior predictive distribution requires a full parameter-space integral at each node in the $(r, n_\text{s})$-plane, and the approximation allows for this to be an efficient calculation. However this assumption could in principle be relaxed. 

The threshold $\bar{\mathcal{E}}= \mathcal{E}_\text{max}/8e$ was chosen such that the predictivities of inflationary models and the inflationary paradigm do not change considerably if we lower this value. In other words we selected this threshold empirically from the criterium that the predictivity values stabilise with this choice; we refer the reader to Appendix \ref{App:PredThreshold} for more details. Notice that, according to the Jeffreys' scale, with this threshold, predictivity is calculated considering events that are inconclusive or only weakly disfavoured compared to the event with maximum evidence $\mathcal{E}_\text{max}$. These events will be what we call events with high evidence. This choice could be changed on a case by case basis.

We finally remark that, as seen in our toy models, different experimental noises result in different predictivities. In Appendix \ref{App:ExNoise} we show how the predictivity changes with the noise for some inflationary models. Fixing the covariance of the sampled data to be given by Eq.~(\ref{CorrMatrix}) corresponds to assuming Planck-like uncertainties for hypothetical experiments resulting in different data. Therefore, the predictivities reported here are to be understood as estimations of testability when observing $r$ and $n_s$ with {\it current} experiments. As the predictivities presented are inherently dependent on the experimental precision, they should be revised for future, lower noise experiments.


\subsection{Results}

We report our results on the selected 85 inflationary models on Table \ref{TablePred} and \ref{TablePred2}. The first column indicates the acronym assigned in \cite{Martin:2013nzq} to the models, the second column shows the Bayes factor with HI as reference model, the third column shows the predictivity of each model, and the fourth column shows the ratio of posterior probabilities after considering predictivities in model priors. 

\begin{figure}[h]
\begin{center}
\scalebox{0.4}{\includegraphics{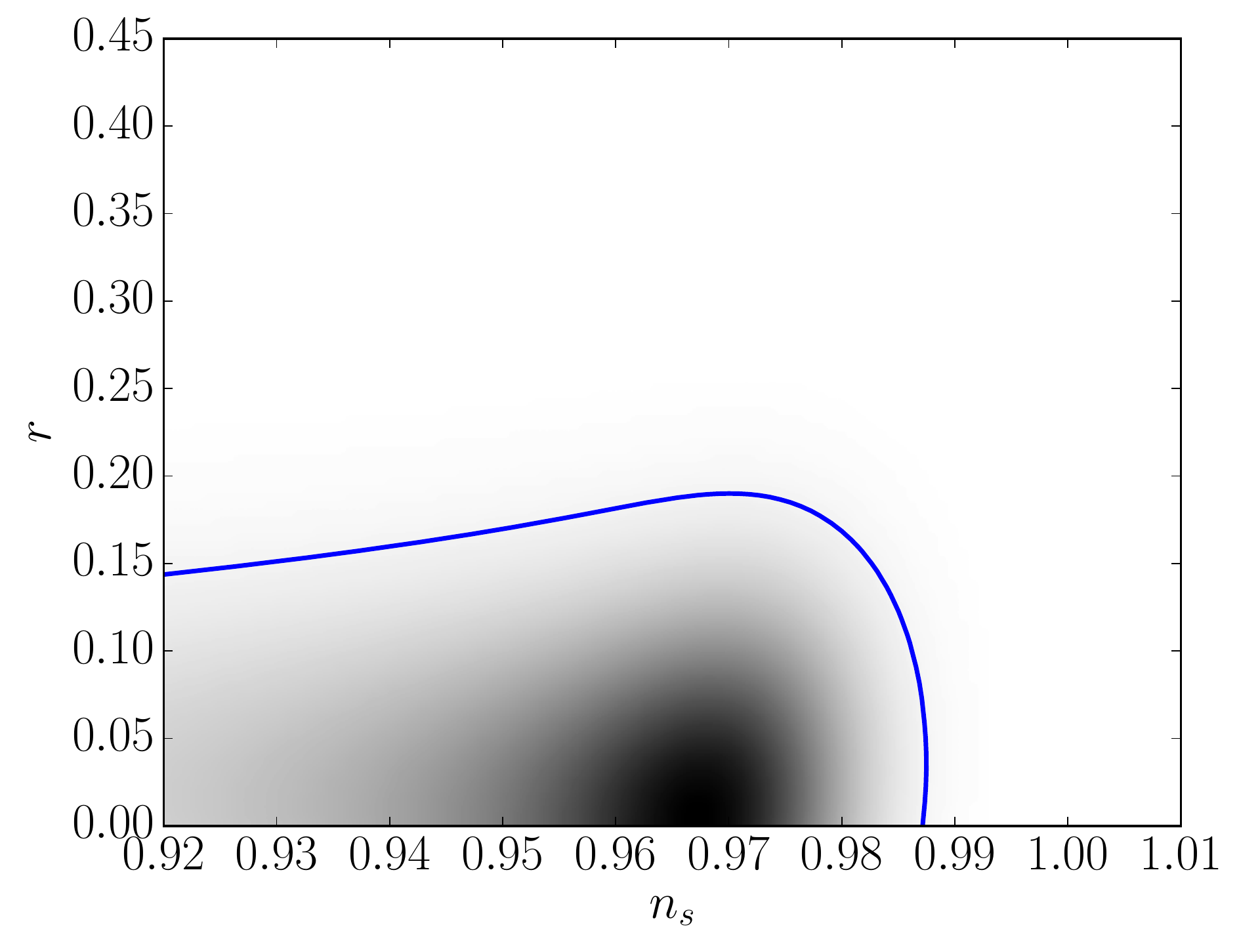}}
\caption{\label{fig:PredHI} Prior predictive distribution for Higgs Inflation (HI). Darker regions have higher evidence, and the blue contour bounds the region with $\mathcal{E} > \mathcal{E}_\text{max}/8e$. From this plot we find a predictivity of $P_r = 0.726$ for HI.}
\end{center}
\end{figure}

For Higgs inflation, we show the prior predictive distribution in Fig.~\ref{fig:PredHI}. Darker regions are those having a larger evidence, and the blue contour bounds the region with a high evidence, i.e~with $\mathcal{E} > \mathcal{E}_\text{max}/8e$. 
As previously explained, the predictivity is calculated as the complement of the area inside the contour over the total area in the plot. We find the predictivity of Higgs inflation to be $P_r=0.726$. Notice that this value already gives information about the model. It means that if we observed any value within the range of the sampled data, there is $73\%$ probability of HI having a low evidence (in the sense of an evidence smaller than $\mathcal{E}_\text{max}/8e$) for that data, and therefore of being disproved. 

\begin{figure}[h]
\begin{center}
\scalebox{0.4}{\includegraphics{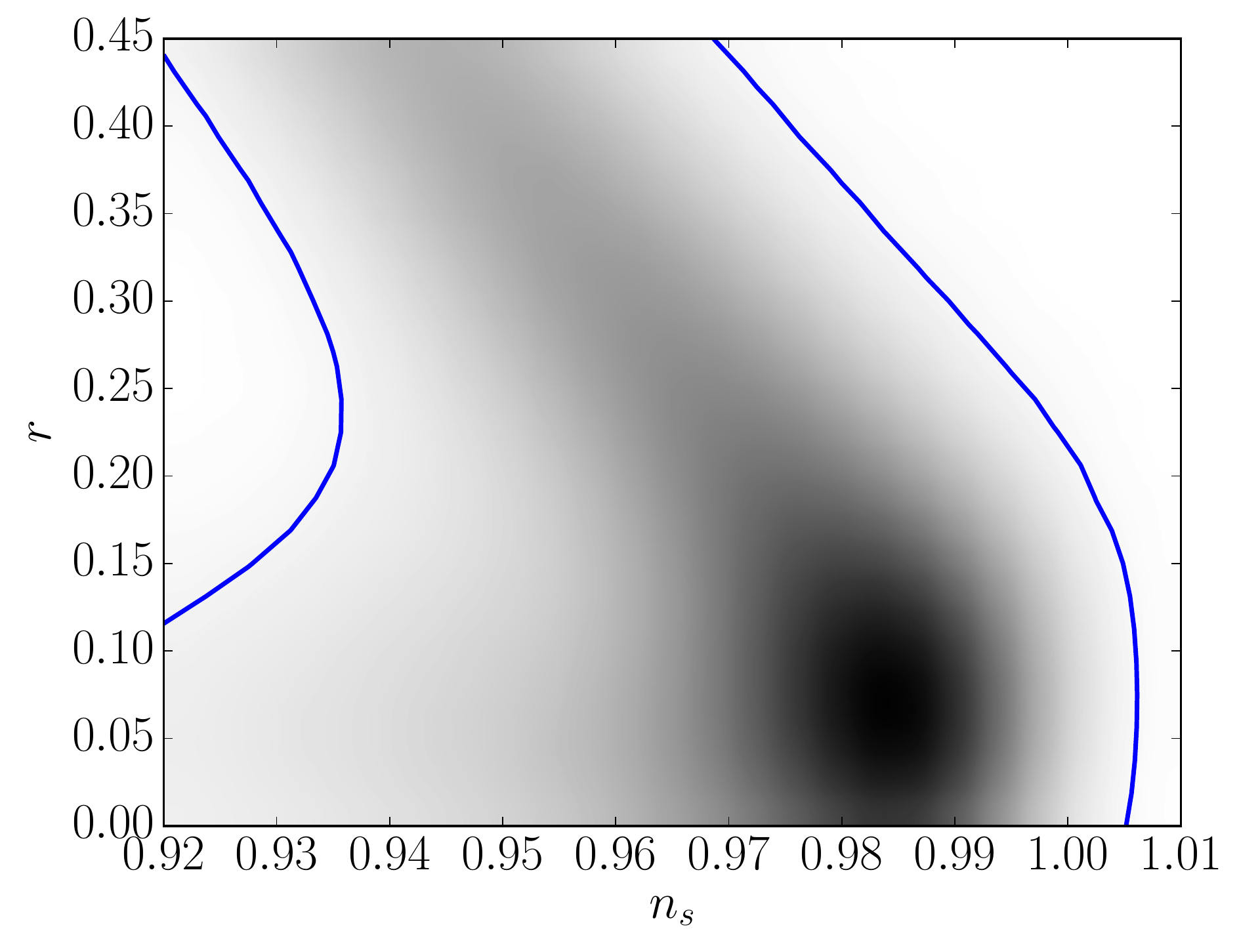}}
\caption{\label{fig:PredRPI3} Prior predictive distribution for RPI3. Darker regions have higher evidence, and the blue contour bounds the region with $\mathcal{E} > \mathcal{E}_\text{max}/8e$. The large area encompassed within this high evidence region corresponds to RPI3 being the least predictive model considered here, with a predictivity of $P_r=0.285$.}
\end{center}
\end{figure}

Fig.~\ref{fig:PredRPI3} shows the corresponding results for RPI3. This is the least predictive model within our set, with $P_r=0.285$. In this case, the Bayes evidence is such that $\ln(\mathcal{E}_{\text{RPI3}}/\mathcal{E}_{\text{HI}})=-2.289$, which, according to Jeffreys' scale, gives a weak evidence of HI over RPI3. We also find that the posterior probability is such that $\ln(P_{\text{RPI3}}/P_{\text{HI}})=-8.44$, which corrects the logarithm of the Bayes factor on $269\%$. This correction changes the conclusion of weak evidence to strong evidence, as unpredictive models are highly penalised.

When comparing the 85 models to HI we find that most of them have comparable predictivities and therefore the incorporation of predictivity into the ratio of posterior probabilities does not change the conclusions on preferences of models. However, there are some exceptions that correspond to very unpredictive models such as: RPI3 ($P_r=0.285$), GMLFI$_{1,3}$ ($P_r=0.362$), LFI ($P_r=0.363$), GMLFI$_{1,2}$ ($P_r=0.382$), LPI1$_{4,1}$ ($P_r=0.414$), among others. In Appendix \ref{App:PredInflModels} we also mention the models with highest predictivities and discuss how they would change with a different sampled data range.

\section{Predictivity of the inflationary paradigm}\label{SecPredInflation}
We are now ready to estimate the predictivity of the inflationary paradigm, and assess how it impacts on its posterior probability. We first estimate the evidence of inflation and compute its Bayes factor relative to an hypothetical paradigm formed by one model only: Higgs inflation (HI). This assumes that information external to cosmology hypothetically selected this model {\it a priori} as the only one viable. We then calculate the predictivity of inflation and incorporate it into the priors to find the ratio of posterior probabilities for these paradigms.

We calculate the evidence of the inflationary paradigm by means of Eq.~(\ref{EvidenceParadigm}), weighting each model with the corresponding prior given by Eq.~(\ref{PriorPr}). Following this procedure we find that the Bayes factor of the inflationary paradigm with respect to that of HI is:
\begin{equation}
\ln(B_{\text{Inf,HI}})=\ln\left(\frac{\mathcal{E}_\text{Inf}}{\mathcal{E}_\text{HI}}\right)=-0.914,
\end{equation}
corresponding to a inconclusive evidence for HI over inflation. This is not altogether surprising, since HI seen as a model is part of the inflationary paradigm, so that its data fitting prowess is included in the inflationary paradigm. The Bayes factor penalises only mildly the fact that the inflationary paradigm also includes models which do not fit the data. 

\begin{figure}[h]
\begin{center}
\scalebox{0.4}{\includegraphics{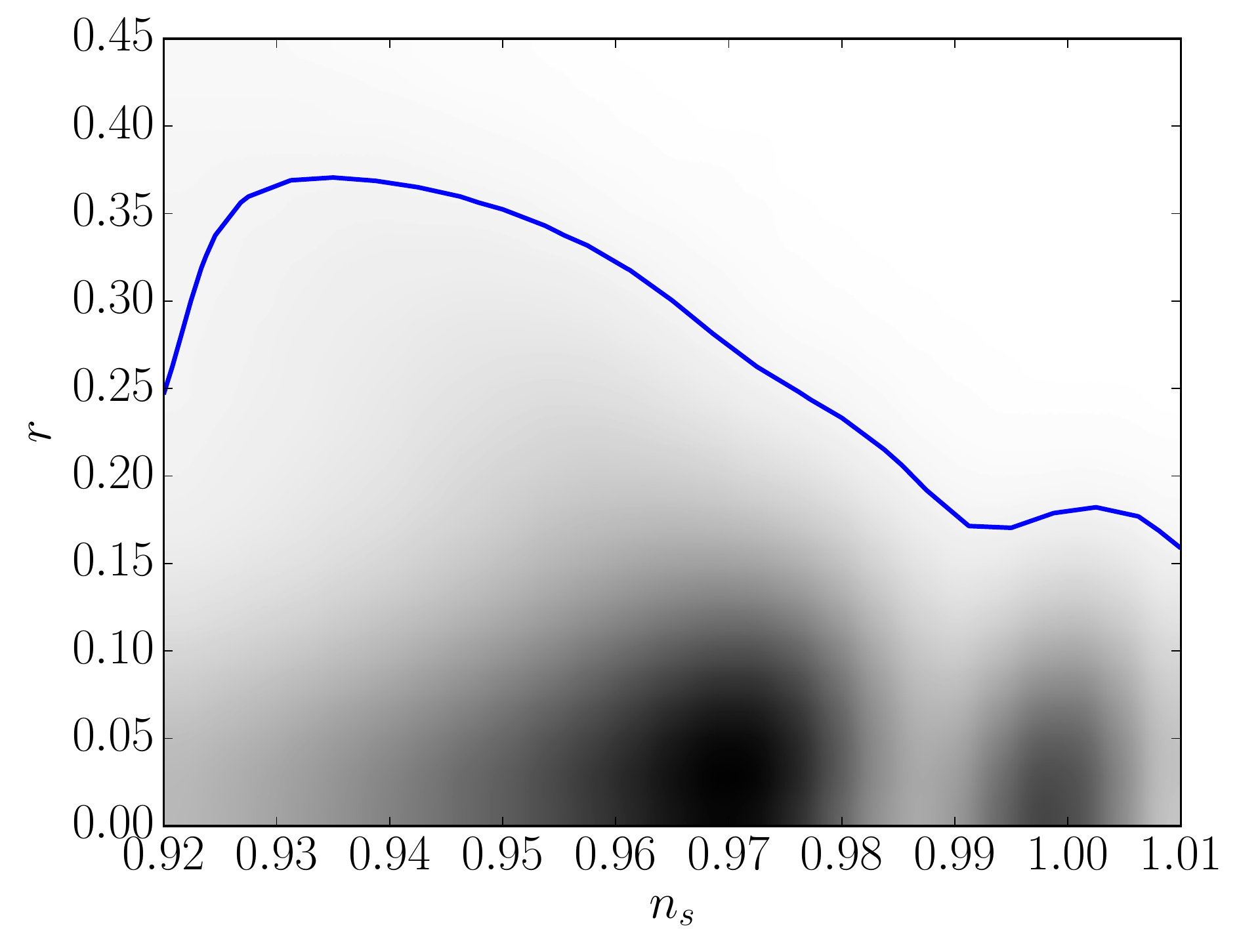}}
\caption{\label{fig:PredInflation} Prior predictive distribution for the inflationary paradigm. Darker regions have higher evidence, and the blue contour bounds the region with $\mathcal{E} > \mathcal{E}_\text{max}/8e$. The predictivity of inflation is found to be $P_r=0.31$.}
\end{center}
\end{figure}

However this changes dramatically when the predictivity is taken into account. In Fig.~\ref{fig:PredInflation} we show how the evidence for the inflationary paradigm changes as the data is varied. This was derived in complete analogy with the procedure followed in the previous section. From this plot we can infer that the predictivity of inflation is $P_r=0.31$. This means that were we to observe any random value within the range of the sampled data, there would be $31\%$ probability of inflation having a low evidence (in the sense of an evidence smaller than $\mathcal{E}_\text{max}/8e$) for that data. This estimates the difficulty in testing the inflationary paradigm.

We incorporate the predictivities of inflation and HI paradigms into their priors according to Eq.~(\ref{PriorParadigm}). This leads to the ratio of posterior probabilities:
\begin{equation}
\ln\left(\frac{P_\text{Inf}}{P_\text{HI}}\right) = -5.707,
\end{equation}
corresponding to a correction of $524\%$ to the logarithm of the Bayes factor. There is a now strong evidence of HI over inflation. As our toy models suggested, the incorporation of predictivity into the Bayesian framework can give considerable changes in the preferences of paradigms.

We should stress that the figure quoted for the predictivity of inflation is a very rough lower bound. We have restricted ourselves to the simplest single-field slow roll models. We considered only one and two parameter models, for numerical reason. We neglected fast roll scenarios. We have also discounted multi-field models (for which the consistency conditions become inequalities). Inclusion of these would degrade inflation's predictivity further and raise the ratio between the posterior of HI over inflation. 

Lowering noise levels would improve the predictivity of inflation, but this would also be true for its competitors, so that its relative status as a scientific theory might not change. One thing that would improve inflation status is the inclusion of information external to cosmology to select {\it a priori} one of its may models, as the HI example shows. We should not forget, however, that {\it what we have presented here is an hypothetical situation}. We do not currently have arguments external to cosmology favouring {\it a priori} HI within all inflationary models. In fact we have selected HI {\it a posteriori}, having seen the data, which is precisely what cannot be done given that almost anywhere in the space of observable there is an inflationary model.


\section{Comparison with an alternative paradigm}\label{VSL}
Let us now examine how inflation fares when compared to alternative paradigms, specifically those making the hard prediction that no primordial gravity waves are produced. There are at least two such possibilities: the cyclic scenario~\cite{Steinhardt:2002ih, Steinhardt:2001st} and the bimetric varying speed of light cosmology~\cite{vsl0,vsl1,vsl2}. We focus on the latter. We remark that bimetric and cyclic scenarios probably fare differently between themselves, because although they make the same prediction this is done at the expense of different sets of parameters (and thus natural priors). 

In bimetric varying speed of light (BVSL) models there are two metrics or ``frames'': metric $g_{\mu\nu}$ which generates the gravitational action (labelled the ``gravity'' metric or the ``Einstein frame''), and metric $\hat g _{\mu\nu}$ to which matter is minimally coupled (the ``matter'' metric or frame). Brans-Dicke (varying $G$) theory has a similar set up, but the two metrics are conformally related. In BVSL models the two metrics are disformally related, according to:
\be
{\hat g}_{\mu\nu}=g_{\mu\nu}+B(\partial_\mu \phi)(\partial_\nu\phi).
\ee
In the simplest scenario the dynamics of $\phi$ is induced by a cosmological constant in the matter frame (possibly balanced by a cosmological constant of opposite sign in the Einstein frame, so as to obtain a trivial low-energy limit). This leads to Klein-Gordon dynamics in the matter frame and (anti-)DBI dynamics in the Einstein frame~\cite{vsl1,moffat}. A power-law potential $V(\phi)$ must then be present and is fully specified by the Bianchi identities, should we require background scaling solutions to exist. If $B$ is constant, $V(\phi)$ should be a mass potential. The cuscuton model\cite{cuscuton} is a limiting case of this model. 

Because the null cones for gravity and for matter are not coincident we have a theory with different speeds of propagation for light (and all massless matter particles) and gravity. If $B>0$ (with signature $(+,-,-,-)$) the speed of light is larger than the speed of gravity. For this reason the model implements the VSL scenario~\cite{VSL-review}. However, whilst the horizon problem is solved for scalar modes, it remains unresolved for tensor modes. This is why the bimetric VSL model (as opposed to DSR realisations~\cite{dsr1,dsr2,dsr3,dsr4}) predicts exactly $r=0$.

In the minimal bimetric model the dimensionful factor $B$ is a constant and exact scale invariance is predicted. However, more general models can be built if we allow $B$ to be a power-law in $\phi$. Then $n_\text{s}\neq 1$ can be accommodated, but $r=0$ remains a solid prediction. The bispectrum is also distorted away from an equilateral shape (associated with $n_\text{s}=1$) and a consistency condition involving $n_\text{s}$ and the bispectrum is found~\cite{vsl1}. This consistency condition contains the bulk of the predictive value of the theory, but it shall be ignored in this paper and left for a further publication. Suffice it to say here that: (1) we are discarding the most predictive part of the theory; (2) the levels of non-Gaussianity generated are consistent with observations~\cite{vsl1}; (3) they could be observable in the future. 

With this proviso in mind, we consider the predictive value of BVSL restricted to observables $n_\text{s}$ and $r$. In such a set up we do not expect 
the model to differ substantially from any sub-class of inflationary models satisfying a linear consistency condition between $n_\text{s}$ and $r$, except for the issue of how the priors upon the model feature. In the Einstein frame the salient features of the model boil down to the one-parameter family of functions for $B$ and $V$:
\bea
B(\phi)&=&\frac{3}{(2-\beta)^2V_0}{\left(\frac{\phi}{M_{p}}\right)}^\beta,\\
V(\phi)&=&V_0{\left(\frac{\phi}{M_{p}}\right)}^{2-\beta}
\eea
where the only relevant parameter is $\beta$, as we shall see. The equation of state associated with scaling solutions is an integration constant, and within a minimal model we should choose radiation ($w=1/3$) to avoid the complications of reheating and phase transitions. With these assumptions we have~\cite{vsl1}:
\bea
n_\text{s}-1&=&\frac{4\beta}{6+\beta},\\
r&=&0.
\eea
The model has other parameters (relevant for fixing, e.g.~the amplitude) but these can be ignored here. 

Just as with inflationary models, a natural choice of prior, failing other information, is uniformity in $\beta$ in the range of the sampled data:
\begin{equation}
\pi(\beta)=U(-0.117,0.015).
\end{equation}
It is interesting to see that a flat prior on $\beta$ translates into a prior for $n_\text{s}$ and $r$ that can be obtained from the inverted function:
\be
\beta=\frac{6(n_\text{s}-1)}{5-n_\text{s}}.
\ee
Thus the prior translates into:
\be
\pi(n_\text{s},r)=\frac{d\beta}{d n_\text{s}}\delta(r).
\ee
For $n_\text{s}\in[0.92,1.01]$ this prior is nearly uniform on $n_\text{s}$ with a peak at the highest $n_s$. In fact, in this range the probability varies by less than $5\%$.

\begin{figure}[h]
\begin{center}
\scalebox{0.4}{\includegraphics{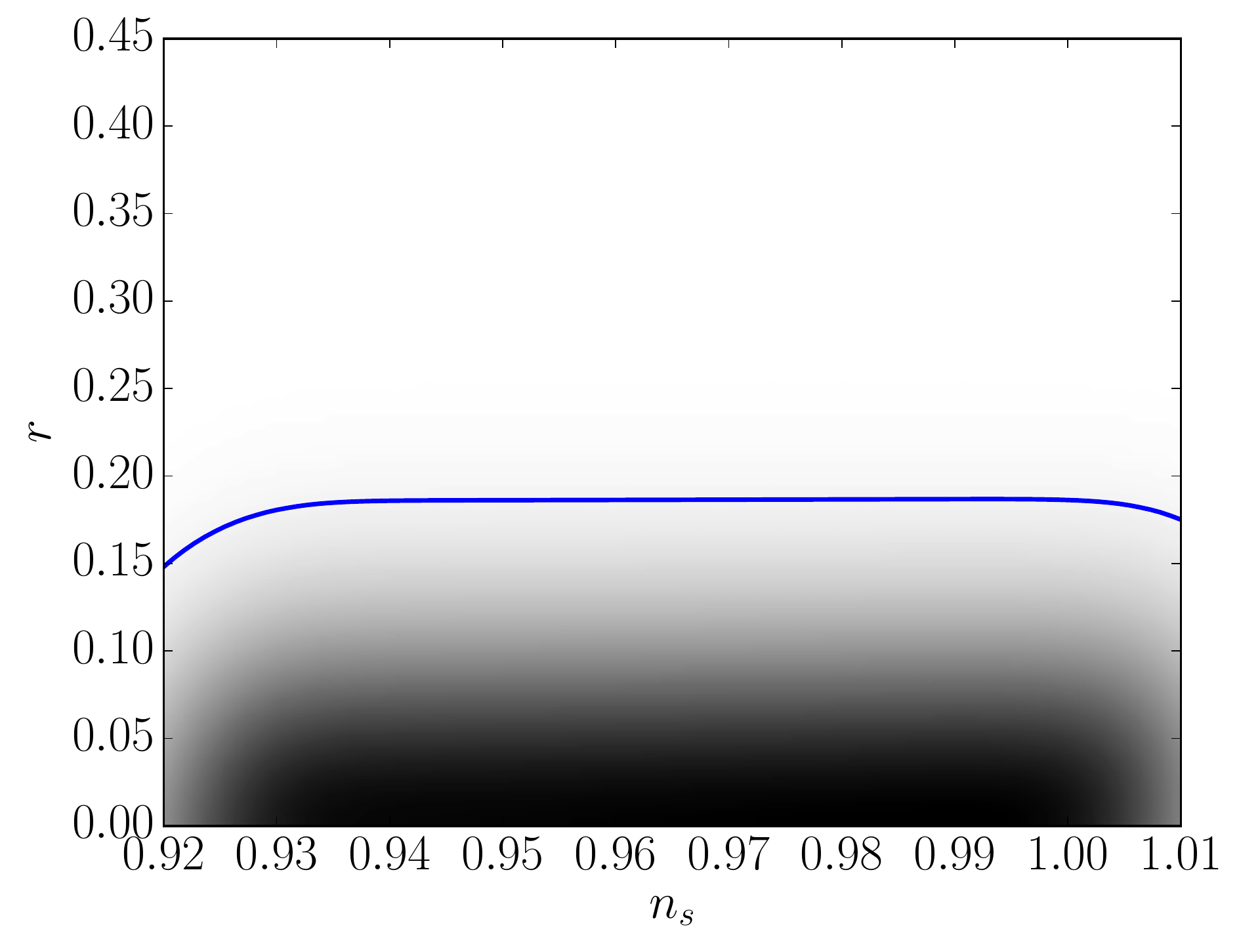}}
\caption{\label{fig:PredBVSL} Prior predictive distribution for BVSL. Darker regions have higher evidence, and the blue contour bounds the region with $\mathcal{E} > \mathcal{E}_\text{max}/8e$. The predictivity of this paradigm is found to be $P_r=0.591$.}
\end{center}
\end{figure}

Figure \ref{fig:PredBVSL} shows the prior predictive distribution of BVSL and the contour bounding the high evidence region ($\mathcal{E} > \mathcal{E}_\text{max}/8e$). From this plot the predictivity of this paradigm is found to be: $P_r=0.591$. When comparing this BVSL paradigm to inflation, the Bayes factor is found to be:
\begin{equation}
\ln(B_{\text{Inf,BVSL}})=\ln\left(\frac{\mathcal{E}_\text{Inf}}{\mathcal{E}_\text{BVSL}}\right)=-0.145,
\end{equation}
which corresponds to an inconclusive evidence for BVSL over inflation. After considering predictivity in the priors of each paradigm we find the ratio of posterior probabilities to be:
\begin{equation}
\ln\left(\frac{P_\text{Inf}}{P_\text{BVSL}}\right) = -4.602,
\end{equation}
which translates into a moderate evidence for BVSL over inflation, according to the Jeffreys' scale. Similarly to the results previously found with HI, there is a large change of $3079\%$ in the logarithm of ratio of probabilities, after introducing predictivity into paradigm comparison.

As in the previous Section, we should stress that this figure is necessarily a very rough lower bound. Not only is the real predictivity of inflation lower than quoted here (as explained in Section~\ref {SecPredInflation}), but the words of caution presented in this Section should be borne in mind: most of the predictive value of non-minimal BVSL involves the bispectrum and this has been ignored here.

\section{Conclusions}\label{SecConclusions}

As the recent hiccups in CMB polarisation observations demonstrated \cite{Ade:2014xna, Ade:2015lrj, Ade:2015tva}, whatever the data turns out to be it will be paraded as proof of inflation. This is because for any observation there is an inflationary model fitting it. Concomitantly, there is a trend in parroting the death of inflation's alternatives (such as cyclic models~\cite{Steinhardt:2002ih, Steinhardt:2001st}, string gas cosmology~\cite{string-gas,string-gas1} and varying speed of light models~\cite{VSL-review,vsl1,dsr1}). 

Putting aside sociology, these perceptions may derive from an important scientific issue. We argued here that the oft-used concept of Bayesian evidence, as applied to paradigms, fails to adequately capture the tenet that
paradigms should be predictive. 
Whilst the standard use of the concept of Bayesian evidence may be perfectly appropriate for comparing models within a paradigm, it fails to suitably penalise the whole paradigm for not making a prediction that could rule it out. 
In this paper we took a first stab at the problem by proposing a measure which we termed ``predictivity'' encoding the view that a predictive theory should not only have a higher evidence than its competitors: its evidence should be {\it exceptional with respect to what it would have been had the data been different}. The proposed ``predictivity'', $P_r$, varies between 0 and 1, and is the percentage of data space producing evidences lower than a given fraction of the maximal evidence. We then proposed priors leading to posteriors such that failure to fit the data and failure to be predictive are both exponentially penalised, putting them on an equal footing.

We applied our formalism to 85 single-field inflationary models, and found that most of them have similar predictivities. Therefore there is not a significant difference between the Bayes factor and the ratio of posterior probabilities when selecting best-fit models within inflation. However, we found that predictivity is crucial in evaluating the success of inflation seen as a whole. We calculated a very rough upper bound based on this collection of models:
\be \label{infl-pred-bound}
P_r < 0.31.
\ee 
This means that had we made a random observation in the range $r\in[0,0.45]$ and $n_\text{s}\in[0.92,1.01]$, there would have been at least a $69\%$ of probability for this observation to have had a high evidence, and thus of favouring inflation. This number estimates the difficulty in ruling out, and thus in testing the inflationary paradigm with present experiments. It also affects drastically the ratios of posteriors between inflation and other paradigms.

We made this point first by considering an hypothetical scenario in which information external to cosmology led to the selection of one of its many models as the only one viable, namely Higgs inflation. {\it We stress that this remains an hypothetical situation}. We do not currently have arguments external to cosmology favouring HI within all inflationary models. In fact we have selected HI {\it a posteriori}, having seen the data, which is precisely what cannot be done and is borderline fraudulent, given that whatever the observations there would have been an inflationary model fitting them. Nonetheless this hypothetical situation allowed us to make an important point: whereas the Bayes factor gives inconclusive evidence of HI seen as a paradigm over inflation seen as a whole, the ratio of posterior probabilities gives strong evidence of HI over inflation. Inclusion of predictivity into the priors corrects the Bayes factor by $524\%$, stressing its importance when evaluating paradigms.

We then compared inflation to the bimetric varying speed of light cosmology (BVSL), when the latter is restricted to its predictions regarding $n_\text{s}$ and $r$. We should not forget that most of the predictive power of BVSL involves the bispectrum, as explained in~\cite{vsl2}, and this has not been factored into our analysis. Nonetheless we can make a point similar to that made for HI regarding the importance of predictivity: this time we found a correction of $3079\%$ in the Bayes factor, after incorporating predictivities into the priors. 
Furthermore, we derived the upper bound on the ratio of the posteriors
\begin{equation}
\ln\left(\frac{P_\text{Inf}}{P_\text{BVSL}}\right) < -4.602,
\end{equation}
revealing at least moderate evidence of BVSL over inflation. This is a lower bound both because the value used for inflation's predictivity
is an upper bound (Eq.~(\ref{infl-pred-bound})) and because we have not used the full predictive power of BVSL.

We stress that Eq.~(\ref{infl-pred-bound}) could be a very rough upper bound indeed, with the real predictivity of inflation much lower. We have restricted ourselves to one and two parameter models for numerical reason, but these are precisely the most predictive single-field inflationary models. We have also discounted multi-field models (for which the consistency conditions become inequalities), and fast-roll models (which cover a larger portion of observable space). Inclusion of these would strongly degrade inflation's predictivity. It would also raise the posterior of HI and BVSL relative to inflation. 
We conclude that cosmic inflation is currently hard to falsify.
However, as the example of HI seen as a paradigm shows, this would change were information external to cosmology select one of its many models as the only one viable.

The inference issues raised in this paper are compounded by the problems concerning the probability and naturalness of inflation. This has been argued in~\cite{stein1,ekpy}, counter-argued in~\cite{counter-stein}, and further argued in~\cite{stein2}. And yet theoretical matters may end up being far more relevant than any probabilistic issues, of whatever nature. The fact that inflation is not an unavoidable part of any quantum gravity framework may prove to be its greatest undoing. Indeed, inflation was historically built to insulate the observable universe from what was felt to be the speculative physics of quantum gravity. By doing so, inflation unwittingly created a situation where such physics is almost certainly forced to remain speculative, if inflation did indeed occur. The quest for quantum gravity continues, with the belief that an observational clue is essential before a viable theory is found. Planck scale physics should be embraced by cosmology, rather than bypassed.

\section*{Acknowledgments}
We thank Erminia Calabrese for useful comments. GG and JM acknowledge support from the John Templeton Foundation. RA is supported by an STFC PhD studentship. ML was funded by Becas Chile. JM was also supported by an STFC consolidated grant and by the Leverhulme Trust.

\appendix
\section{Probability and predictivity with double-valuation}\label{App:doublevalue}

\subsection{Probabilty with double-valuation}\label{App:ProbDouble}
For a given model we want to calculate the probability of the evidence $P(\mathcal{E})$ within a range of sampled data $\mathtt{d}\in [\mathtt d_1,\mathtt d_2]$. Within this range there is going to be a maximum ${\cal E}_\text{max}$ and minimum ${\cal E}_\text{min}$ of the evidence. The probability of the evidence will be such that $P(\mathcal{E})=0$ for $\mathcal{E}>\mathcal{E}_\text{max}$ and $\mathcal{E}<\mathcal{E}_\text{min}$.

On the other hand, in general, the evidence as a function of the data will be a 1-peaked double valued function $\mathcal{E}(\mathtt d)$ in the range of the sampled data \footnote{The arguments appearing in this section can also be straightforwardly generalised to the case of a multi-peaked evidence}. In this case the evidence can be broken up into two 1-1 functions around a value $\theta_\text{max}$:
\begin{equation}
\mathcal{E}(\mathtt d)= \left\{ \begin{array}{ll}
\mathcal{E}_1(\mathtt d) & : {\rm if} \quad \mathtt d_1<\mathtt d<\theta_\text{max}\\
 \mathcal{E}_2(\mathtt d) & :{\rm if} \quad \theta_\text{max}<\mathtt d<\mathtt d_2
\end{array}
\right. , 
\end{equation}
where $\theta_\text{max}$ is such that $\mathcal{E}(\theta_\text{max})=\mathcal{E}_\text{max}$. 
In each piece we define the probabilities $P_1$ and $P_2$ such that: 
\begin{equation}
P_1( \mathcal{E}_1)|d\mathcal{E}_1| =P_1(\mathtt d)|d\mathtt d| ; \quad P_2( \mathcal{E}_2)|d\mathcal{E}_2| =P_2(\mathtt d)|d\mathtt d|,
\end{equation}
and therefore the probabilities $P_1$ and $P_2$ can be calculated separately as:
\begin{equation}
P_1(\mathcal{E})=P_1(\mathtt d)\biggr|\frac{d\mathcal{E}_1(\mathtt d)}{d\mathtt d}\biggr|^{-1};\quad P_2(\mathcal{E})=P_2(\mathtt d)\biggr|\frac{d\mathcal{E}_2(\mathtt d)}{d\mathtt d}\biggr|^{-1}.
\end{equation}
The total probability will be:
\begin{equation}
P(\mathcal{E})=P_1(\mathcal{E})+P_2(\mathcal{E}).
\end{equation}

\subsection{Predictivity with double-valuation}\label{App:PredDouble}

The predictivity of the model will be:
\begin{equation}
P_r=1-\int_{\bar{\mathcal{E}}}^{\mathcal{E}_\text{max}}P(\mathcal{E})d\mathcal{E},
\end{equation}
but this integral can be re-written as:
\begin{align}\label{App:IntegralProbability}
&\int_{\bar{\mathcal{E}}}^{\mathcal{E}_\text{max}}P(\mathcal{E})d\mathcal{E}=\int_{\bar{\mathcal{E}}}^{\mathcal{E}_\text{max}}P_1(\mathcal{E})d\mathcal{E}+\int_{\bar{\mathcal{E}}}^{\mathcal{E}_\text{max}}P_2(\mathcal{E})d\mathcal{E}\nonumber\\
& =\int_{\mathtt{d}_1}^{\theta_\text{max}}P_1(\mathtt d)|d\mathtt d|+\int_{\mathtt{d}_2}^{\theta_\text{max}}P_2(\mathtt d)|d\mathtt d|\nonumber\\
& =\int_{\mathtt{d}_1}^{\theta_\text{max}}P_1(\mathtt d)d\mathtt d+\int^{\mathtt{d}_2}_{\theta_\text{max}}P_2(\mathtt d)d\mathtt d,\nonumber \\
& =\int_{-\infty}^{\infty}P_1(\mathtt d)H(\mathcal{E}(\mathtt d)-\bar{\mathcal{E}})d\mathtt d+\int^{\infty}_{-\infty}P_2(\mathtt d)H(\mathcal{E}(\mathtt d)-\bar{\mathcal{E}}) d\mathtt d,
\end{align}
where $\bar{\mathcal{E}}=\mathcal{E}(\mathtt{d}_1)=\mathcal{E}(\mathtt{d}_2)$, and $H(x)$ is the Heaviside step function: $H(x)=1$ if $x>0$, and $H(x)=0$ if $x<0$. For a flat prior on the sampled data, we use a similar expression to that given by Eq.~(\ref{FlatData}):
\begin{align}
&P_1(\mathtt d)=\frac{H(\theta_\text{max}-\mathtt d)-H(\mathtt d_1-\mathtt d)}{\Delta \mathtt d} ,\\
& P_2(\mathtt d)=\frac{H(\mathtt d_2-\mathtt d)-H(\theta_\text{max}-\mathtt d)}{\Delta \mathtt d},
\end{align}
where $\Delta \mathtt d= \mathtt d_2-\mathtt d_1$. Using these equations, Eq.~(\ref{App:IntegralProbability}) becomes:
\begin{align}
& = \int_{-\infty}^{\infty}\frac{1}{\Delta \mathtt d}\left(H(\theta_\text{max}-\mathtt d)-H(\mathtt d_1-\mathtt d)\right) H(\mathcal{E}(\mathtt d)-\bar{\mathcal{E}}) d\mathtt d \nonumber \\
& + \int^{\infty}_{-\infty}\frac{1}{\Delta \mathtt d}\left(H(\mathtt d_2-\mathtt d)-H(\theta_\text{max}-\mathtt d)\right) H(\mathcal{E}(\mathtt d)-\bar{\mathcal{E}}) d\mathtt d\nonumber \\
& = \frac{1}{\Delta \mathtt d}\int_{\mathtt d_1}^{\theta_\text{max}} H(\mathcal{E}(\mathtt d)-\bar{\mathcal{E}}) d\mathtt d + \frac{1}{\Delta \mathtt d} \int^{\mathtt d_2}_{\theta_\text{max}}H(\mathcal{E}(\mathtt d)-\bar{\mathcal{E}}) d\mathtt d \nonumber\\
& = \frac{1}{\Delta \mathtt d} \int^{\mathtt d_2}_{\mathtt d_1}H(\mathcal{E}(\mathtt d)-\bar{\mathcal{E}}) d\mathtt d.
\end{align}

Therefore, the predictivity can be written as:
\begin{equation}\label{PredHeaviside}
P_r=1-\frac{1}{\Delta \mathtt d}\int_{\mathtt d_1}^{\mathtt d_2}H(\mathcal{E}(\mathtt d)-\bar{\mathcal{E}})d\mathtt d.
\end{equation}

We conclude that the predictivity is going to be the complement of the ratio of the range of data ({\it within} the range of the sampled data) giving an evidence larger than $\bar{\mathcal{E}}$ over the whole sampled data range.

\section{Probability and predictivity in $N$ dimensions}\label{App:PredNdim}

In general, let us consider a model such that its evidence is a function of $N$ observables $\mathcal{E}(\theta_1,...,\theta_N)$. The probability density function of $\mathcal{E}$ can be written as:
\begin{equation}\label{App:MultiProb}
P(\mathcal{E})=\int_{V_s} d\theta_1...d\theta_N \; P(\theta_1)...P(\theta_N)\delta(\mathcal{E}-\mathcal{E}(\theta_1,...,\theta_N)),
\end{equation}
where the range of the integrals corresponds to the range of the sampled data, denoted generically by an $N$-volume $V_s$, because outside this range $P(\theta_i)=0$. Notice that this formula is always true no matter if there are multiple valuations in the function $\mathcal{E}(\theta_1,...,\theta_N)$.

The predictivity of the model will be:
\begin{equation}
P_r=1-\int_{\bar{\mathcal{E}}}^{\mathcal{E}_\text{max}}P(\mathcal{E})d\mathcal{E},
\end{equation}
but using Eq.~(\ref{App:MultiProb}) we can rewrite this last integral as:
\begin{equation}\label{PredNdim1}
\int_{V_s} d\theta_1...d\theta_N \; P(\theta_1)...P(\theta_N) \int_{\bar{\mathcal{E}}}^{\mathcal{E}_\text{max}}\delta(\mathcal{E}-\mathcal{E}(\theta_1,...,\theta_N)) d\mathcal{E}.
\end{equation}
On the other hand, this last integral in $d\mathcal{E}$ can be written as:
 \begin{align}
&\int_{\bar{\mathcal{E}}}^{\mathcal{E}_\text{max}}\delta(\mathcal{E}-\mathcal{E}(\theta_1,...,\theta_N)) d\mathcal{E}\nonumber\\
& =H(\mathcal{E}_\text{max}-\mathcal{E}(\theta_1,...,\theta_N))-H(\mathcal{E}(\theta_1,...,\theta_N)-\bar{\mathcal{E}})\nonumber\\
&= 1-H(\mathcal{E}(\theta_1,...,\theta_N)-\bar{\mathcal{E}})\nonumber\\
&=H(\bar{\mathcal{E}}-\mathcal{E}(\theta_1,...,\theta_N)),
\end{align}
where we have used that by construction, $\mathcal{E}(\theta_1,...,\theta_N) \leq \mathcal{E}_\text{max}$.
Then, Eq.~(\ref{PredNdim1}) becomes:
\begin{align}
&\int_{V_s} d\theta_1...d\theta_N \; P(\theta_1)...P(\theta_N) H(\bar{\mathcal{E}}-\mathcal{E}(\theta_1,...,\theta_N)),\nonumber\\
& =\frac{1}{\Delta\theta_1}...\frac{1}{\Delta\theta_N}\int_{V_s} d\theta_1...d\theta_N \; H(\bar{\mathcal{E}}-\mathcal{E}(\theta_1,...,\theta_N)),
\end{align}
which is the same result that we obtained for the 1-dimensional case. We conclude that the predictivity is always going to be the complement of the ratio of the N-volume of data ({\it within} the N-volume of the sampled data $V_s$) giving an evidence larger than $\bar{\mathcal{E}}$ over the whole N-volume sampled data.

\section{Predictivity of inflationary models}\label{App:PredInflModels}
In this section we report the results on predictivities for 85 inflationary models described in \cite{Martin:2014vha}. 

Table \ref{TablePred} continues in Table \ref{TablePred2}. Tables \ref{TablePred} and \ref{TablePred2} show the ratios of evidences when using Higgs inflation as the reference model (second column), the predictivity for each model (third column), and ratios of posterior probabilities as in Eq.~(\ref{ratio-model}) when the prior for each model is considered to be given by Eq.~(\ref{PriorPr}) (fourth column). The names of models are in the first column of the tables, and correspond to the initials given in \cite{Martin:2014vha}.

\begin{table}[h!]
\scriptsize
\begin{multicols}{2}
\centering
\begin{tabular}{| c || c || c || c |}
\hline
Model & $\ln(\mathcal{E}/\mathcal{E}_{\text{HI}})$ & $Pr$ & $\ln(P/P_\text{HI}) $ \\ \hline
AI & -0.172 & 0.726 & -0.171\\ \hline
$\text{BI}_{1\text{s}}$ & -0.194 & 0.715 & -0.21\\ \hline
BI$_{2\text{s}}$ & -0.073 & 0.73 & -0.067\\ \hline
BI$_{3\text{s}}$ & -0.021 & 0.728 & -0.018\\ \hline
BI$_{4\text{s}}$ & 0.006 & 0.738 & 0.022\\ \hline
BI$_{5\text{s}}$ & 0.021 & 0.736 & 0.035\\ \hline
BI$_{6\text{s}}$ & 0.031 & 0.736 & 0.045\\ \hline
CNAI & -2.177 & 0.44 & -3.654\\ \hline
CNBI & -1.888 & 0.499 & -2.752\\ \hline
CWI$_\text{f}$ & -1.373 & 0.782 & -1.307 \\ \hline
CWI$_\text{l}$ & -1.365 & 0.781 & -1.301 \\ \hline
DWI & -2.563 & 0.432 & -4.149 \\ \hline
ESI & 0.027 & 0.725 & 0.026\\ \hline
ESI$_\text{l}$ & -0.531 & 0.57 & -0.959\\ \hline
ESI$_\text{o}$ & 0.043 & 0.725 & 0.041\\ \hline
ESI$_{\sqrt{2}}$ & 0.034 & 0.738 & 0.05\\ \hline
ESI$_{\sqrt{2/3}}$ & 0.018 & 0.725 & 0.017\\ \hline
\end{tabular}
\centering
\begin{tabular}{| c || c || c || c |}
\hline
Model & $\ln(\mathcal{E}/\mathcal{E}_{\text{HI}})$ & $Pr$ & $\ln(P/P_\text{HI}) $ \\ \hline
GMLFI$_{2/3,1/3}$ & -1.39 & 0.565 & -1.841\\ \hline
GMLFI$_{2/3,4/3}$ & -1.871 & 0.45 & -3.227\\ \hline
GMLFI$_{1,1}$ & -2.092 & 0.486 & -3.064\\ \hline
GMLFI$_{1,2}$ & -2.811 & 0.382 & -5.277\\ \hline
GMLFI$_{1,3}$ & -4.068 & 0.362 & -7.042\\ \hline
GMLFI$_{2,1}$ & -3.518 & 0.472 & -4.627\\ \hline
GMLFI$_{2,3}$ & -6.039 & 0.522 & -6.738\\ \hline
GMLFI$_{3,1}$ & -5.649 & 0.515 & -6.392\\ \hline
GMLFI$_{3,2}$ & -6.523 & 0.539 & -7.11\\ \hline
GMLFI$_{3,3}$ & -9.102 & 0.632 & -9.298\\ \hline
HF1I & -2.997 & 0.493 & -3.914\\ \hline
HI & 0 & 0.726 & 0\\ \hline
IMI1 & -13.268 & 0.899 & -13.138 \\ \hline
IMI2 & -13.247 & 0.891 & -13.119 \\ \hline
IMI3 & -13.096 & 0.891 & -12.968 \\ \hline
IMI4 & -12.725 & 0.894 & -12.597\\ \hline
IMI5 & -12.572 & 0.896 & -12.442 \\ \hline
\end{tabular}
\end{multicols}
\caption{\label{TablePred} Values for ratio of evidences, predictivities, and ratio of posterior probabilities for 85 single-field slow-roll inflationary models. Results continue on Table \ref{TablePred2}.}
\end{table}

\FloatBarrier

\begin{table}[h]
\scriptsize
\begin{multicols}{2}
\centering
\begin{tabular}{| c || c || c || c |}
\hline
Model & $\ln(\mathcal{E}/\mathcal{E}_{\text{HI}})$ & $Pr$ & $\ln(P/P_\text{HI}) $ \\ \hline
IMI6 & -12.769 & 0.896 & -12.64 \\ \hline
KMII & 0.043 & 0.728 & 0.046\\ \hline
KMII$_{V>0}$ & 0.023 & 0.73 & 0.028\\ \hline
LFI & -2.491 & 0.363 & -5.423\\ \hline
LFI$_1$ & -1.718 & 0.546 & -2.269\\ \hline
LFI$_2$ & -3.026 & 0.493 & -3.943\\ \hline
LFI$_3$ & -5.021 & 0.507 & -5.822\\ \hline
LFI$_4$ & -8.027 & 0.574 & -8.433\\ \hline
LFI$_{2/3}$ & -1.317 & 0.579 & -1.702\\ \hline
LI & -0.259 & 0.554 & -0.767\\ \hline
LI$_{\alpha<0}$ & -1.896 & 0.531 & -2.532\\ \hline
LI$_{\alpha>0}$ & -0.476 & 0.738 & -0.46\\ \hline
LPI1$_{4,1}$ & -3.595 & 0.414 & -5.449\\ \hline
LPI1$_{4,2}$ & -5.032 & 0.496 & -5.921\\ \hline
LPI1$_{4,3}$ & -7.181 & 0.563 & -7.639\\ \hline
MHI & -0.292 & 0.541 & -0.87\\ \hline
MHI$_\text{l}$ & 0.048 & 0.742 & 0.07\\ \hline
MHI$_\text{s}$ & -0.789 & 0.48 & -1.82\\ \hline
MLFI & -4.309 & 0.458 & -5.571\\ \hline
MSSMI$_\text{o}$ & -4.607 & 0.824 & -4.51\\ \hline
MSSMI$_\text{p}$ & -3.87 & 0.464 & -5.062\\ \hline
NI & -2.608 & 0.43 & -4.217\\ \hline
OSTI & -2.123 & 0.456 & -3.404\\ \hline
PLI & -8.472 & 0.856 & -8.357\\ \hline
PLI$_\text{p}$ & -8.373 & 0.47 & -9.497 \\ \hline
RCHI & -1.611 & 0.621 & -1.842\\ \hline
\end{tabular}
\centering
\begin{tabular}{| c || c || c || c |}
\hline
Model & $\ln(\mathcal{E}/\mathcal{E}_{\text{HI}})$ & $Pr$ & $\ln(P/P_\text{HI}) $ \\ \hline
RCHI$_o$ & -88.064 & 0.976 & -87.922\\ \hline
RCMI & -2.673 & 0.434 & -4.237\\ \hline
RCQI & -6.692 & 0.424 & -8.395\\ \hline
RGI & -0.365 & 0.565 & -0.816\\ \hline
RGI$_{1/16}$ & -0.096 & 0.725 & -0.098\\ \hline
RGI$_\text{l}$ & -0.758 & 0.501 & -1.609\\ \hline
RGI$_\text{s}$ & -0.075 & 0.725 & -0.076\\ \hline
RIPI$_\text{o}$ & -5.038 & 0.83 & -4.937\\ \hline
RIPI$_\text{p}$ & -2.842 & 0.4.88 & -3.782\\ \hline
RIPI$_\text{sugra}$ & -1.128 & 0.53 & -1.774\\ \hline
RPI1 & -3.619 & 0.714 & -3.64\\ \hline
RPI3 & -2.289 & 0.285 & -8.453\\ \hline
SBI$_{\alpha_{\text{min}}}$ & -1.703 & 0.778 & -1.642\\ \hline
SFI$_1$ & -2.404 & 0.533 & -3.03\\ \hline
SFI$_2$ & -1.974 & 0.554 & -2.482\\ \hline
SFI$_{2\text{l}}$ & -1.574 & 0.554 & -2.081\\ \hline
SFI$_3$ & -1.515 & 0.611 & -1.777\\ \hline
SFI$_{3\text{l}}$ & -1.119 & 0.584 & -1.484\\ \hline
SFI$_{3\text{s}}$ & -5.583 & 0.83 & -5.482\\ \hline
SFI$_4$ & -0.784 & 0.648 & -0.936\\ \hline
SFI$_{4\text{l}}$ & -0.615 & 0.624 & -0.836\\ \hline
SFI$_{4s}$ & -1.239 & 0.778 & -1.178\\ \hline
TI$_{1/2}$ & -1.255 & 0.778 & -1.194\\ \hline
WRI$_\text{g}$ & -1.252 & 0.467 & -2.41\\ \hline
WRI$_\text{o}$ & -1.179 & 0.587 & -1.531\\ \hline
\end{tabular}
\end{multicols}
\caption{\label{TablePred2} Values for ratio of evidences, predictivities, and ratio of posterior probabilities for 85 single-field slow-roll inflationary models. This table is a continuation of Table \ref{TablePred}. We note that for numerical reasons the prior of model SFI$_1$ was set to: $\pi(\log(\mu/M_p))=U(0,2)$.}
\end{table}

As we can see in the tables, most of the models have a similar predictivity, and therefore the preferences for models do not change significantly after considering predictivity. There are some exceptions of very unpredictive models such as: RPI3 ($P_r=0.285$), GMLFI$_{1,3}$ ($P_r=0.362$), LFI ($P_r=0.363$), GMLFI$_{1,2}$ ($P_r=0.382$), LPI1$_{4,1}$ ($P_r=0.414$), among others. 

\begin{figure}[h]
\begin{center}
\scalebox{0.4}{\includegraphics{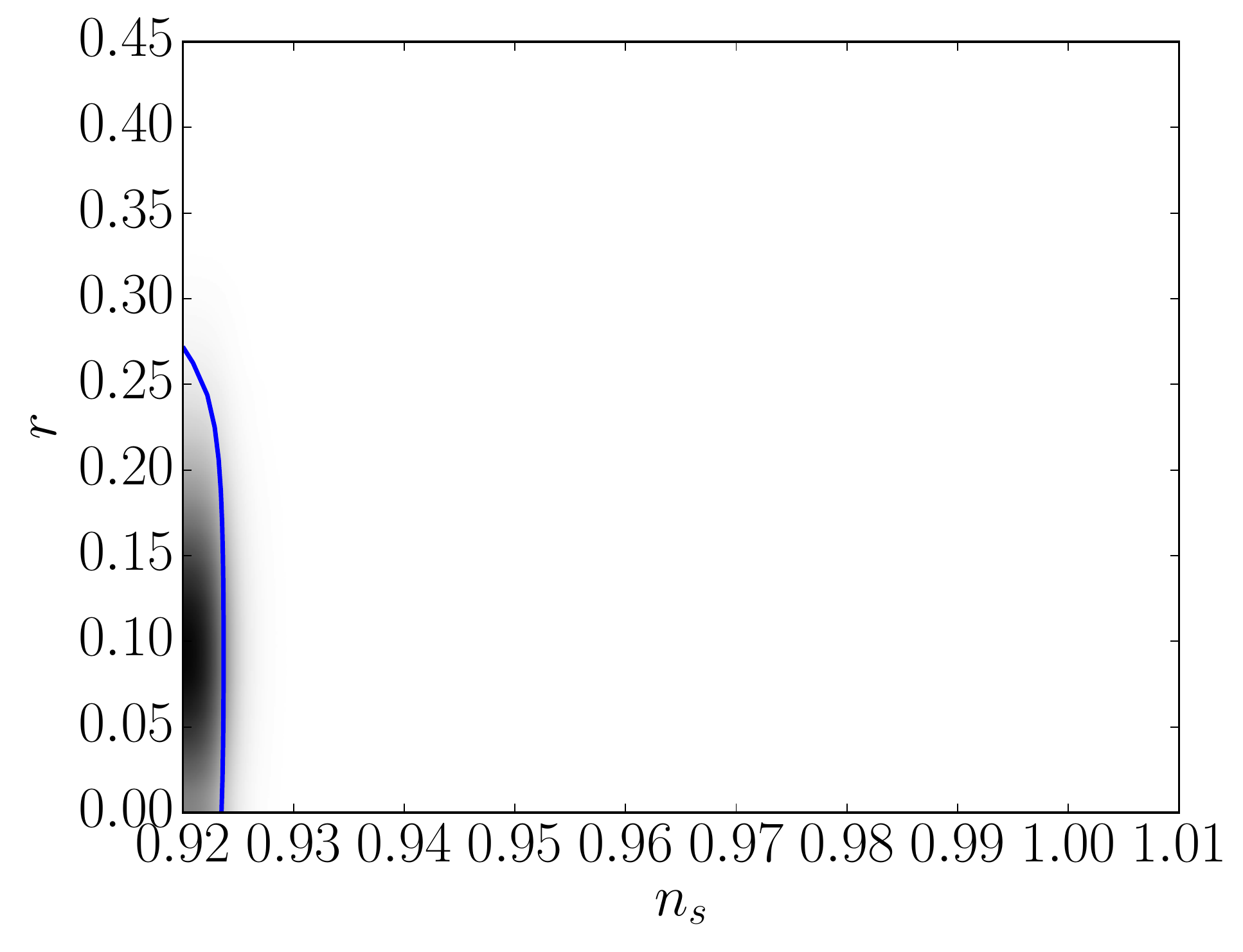}}
\caption{\label{fig:PredRCHIo} Prior predictive distribution for RCHI$_\text{o}$. Darker regions have higher evidence, and the blue contour bounds the region with $\mathcal{E} > \mathcal{E}_\text{max}/8e$. This model is the one with the highest predictivity, with $P_r = 0.976$.}
\end{center}
\end{figure}

The least predictive model is RPI3 with $P_r=0.285$, and its prior predictive distribution is shown in Fig.~\ref{fig:PredRPI3}. The most predictive model is RCHI$_\text{o}$ with $P_r=0.976$. Model RCHI$_\text{o}$ is defined by the radiatively corrected Higgs inflationary potential:
\begin{equation}
V(\phi)=M^4\left(1-2e^{-(2/\sqrt{6})(\phi/M_p)}+\frac{A_1}{16\pi^2}\frac{\phi}{\sqrt{6}M_p}\right),
\end{equation}
where $A_1$ is the only free relevant potential parameter. The prior choice for $A_1$ is a flat prior $\pi(A_1)=U(-48,-20) $. As it can be seen from Table \ref{TablePred2}, RCHI$_\text{o}$ is the most predictive model but it is highly disfavoured by data, as it has a low evidence compared to HI. Fig.~\ref{fig:PredRCHIo} shows the prior predictive distribution for this model. 

\begin{figure}[h!]
\begin{center}
\scalebox{0.4}{\includegraphics{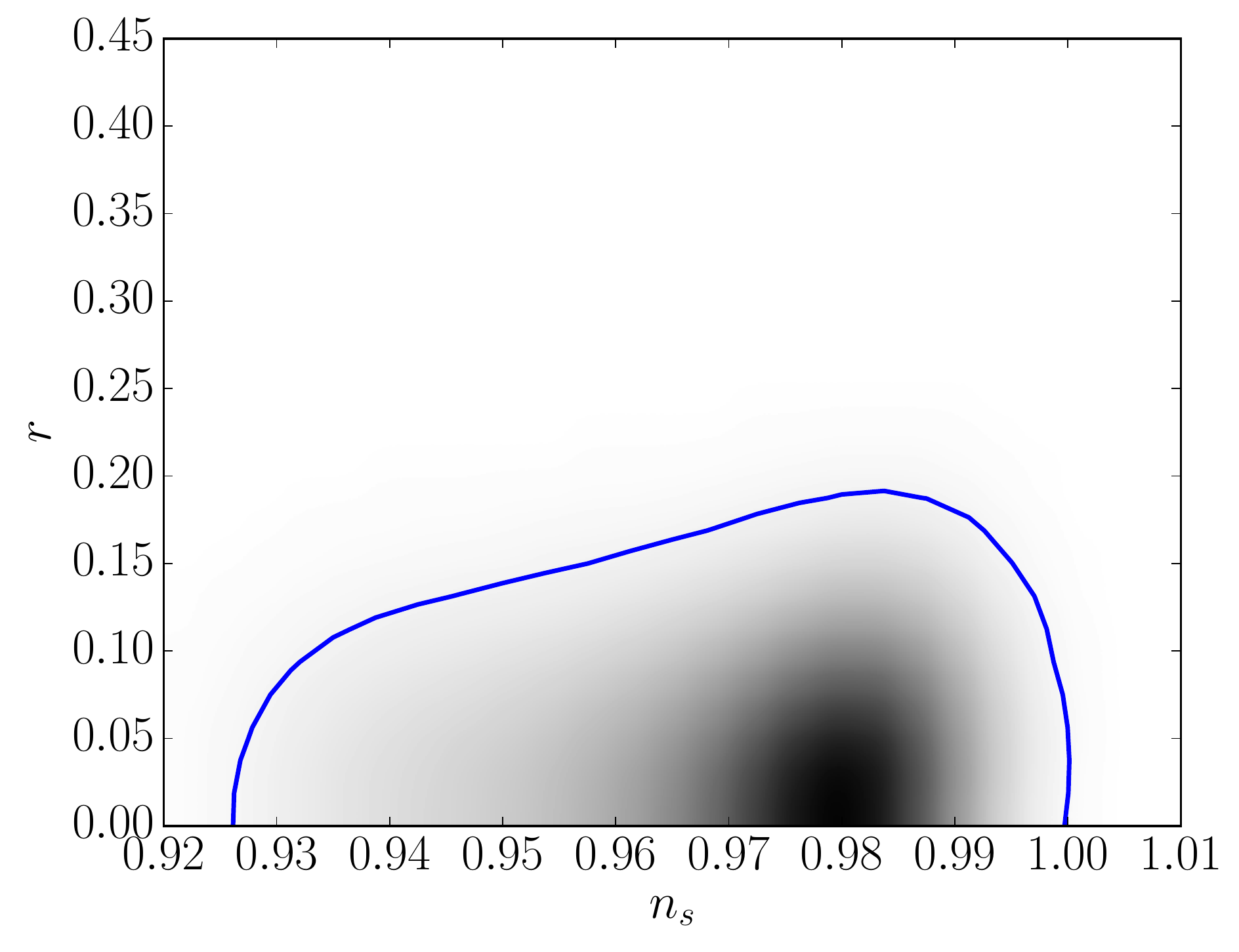}}
\caption{\label{fig:PredLIapos} Prior predictive distribution for LI$_{\alpha>0}$. Darker regions have higher evidence, and the blue contour bounds the region with $\mathcal{E} > \mathcal{E}_\text{max}/8e$. The predictivity of this model is $P_r = 0.738$.}
\end{center}
\end{figure}

From this plot we conclude two things: first, this model does not fit well the data as it favours low values for $n_\text{s}$ compared to the best-fit from Planck. Second, it is clear that the high predictivity is due to the choice on the sampled data, and that a larger data range might give a lower relative predictivity, as the region contained inside the blue contour is expected to grow for this model, but not for other models. For this reason we also mention another model with one of the highest predictivities, LI$_{\alpha>0}$ with $P_r=0.738$. This model is described by the loop inflationary potential:
\begin{equation}
V(\phi)=M^4\left[1+\alpha\ln\left(\frac{\phi}{M_p}\right)\right],
\end{equation}
where $\alpha>0$ is the only free relevant potential parameter. The prior defining this model is given by $\pi(\log(\alpha))=U(\log(3\times 10^{-3} ),\log(3\times 10^{-1}))$. Fig.~\ref{fig:PredLIapos} shows the prior predictive distribution for this model, and the contour bounding the high evidence region. Contrary to RCHI$_\text{o}$, from Fig.~\ref{fig:PredLIapos} we can see that the region inside the blue contour would not change with a larger sampled data (assuming $r>0$). 

\FloatBarrier

\section{Predictivity threshold}\label{App:PredThreshold}

In Eq.~(\ref{Pred}) we have to assume a threshold $\bar{\mathcal{E}}$ for evaluating the predictivity. As previously mentioned, this value is arbitrary, and in this section we make an analysis for finding an appropriate value for this threshold. 

\begin{figure}[h]
\begin{center}
\scalebox{0.4}{\includegraphics{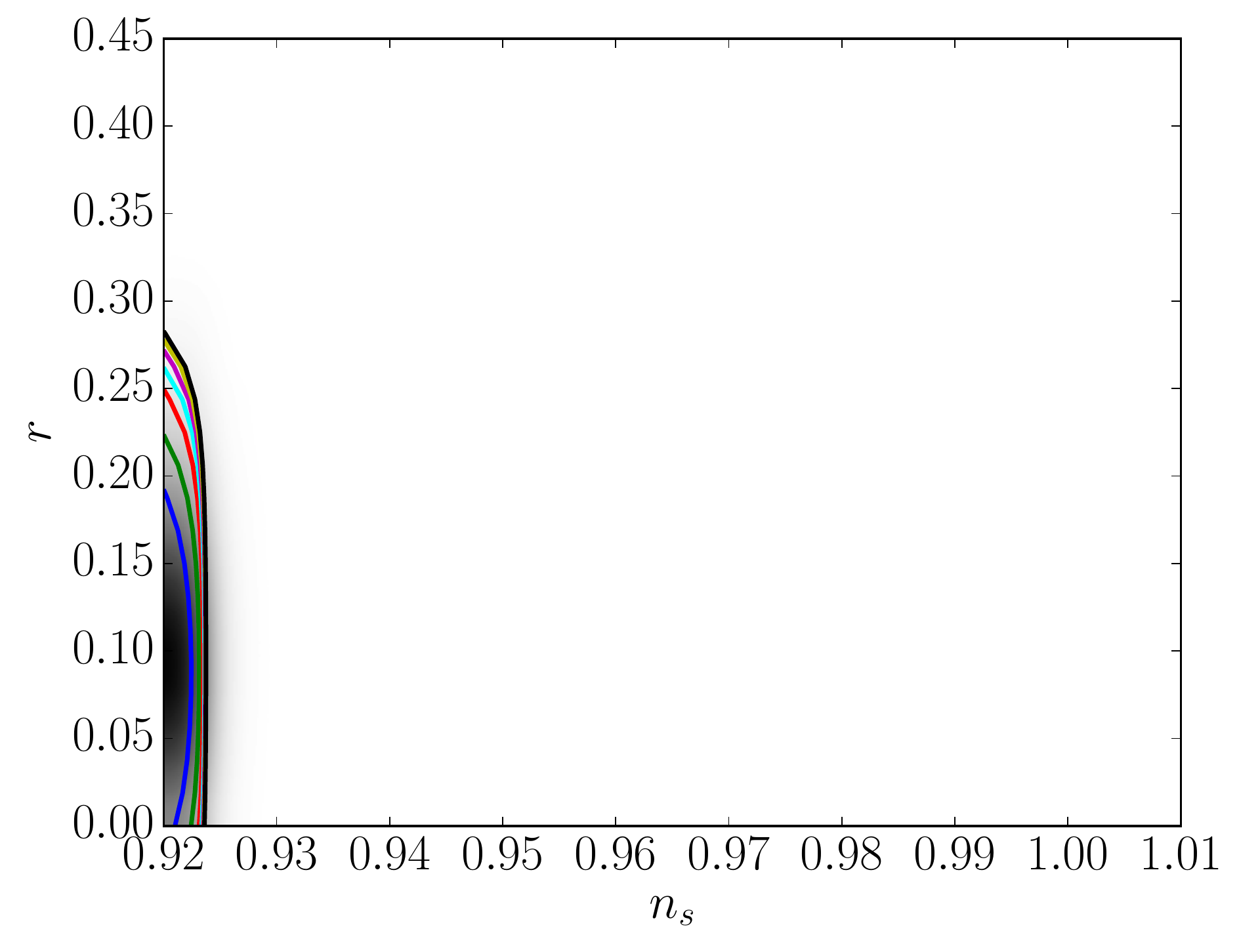}}
\caption{\label{fig:RCHIoContours} Prior predictive distribution for RCHIo. Darker regions have higher evidence. We also show the contours bounding regions with evidence higher than $\mathcal{E}_\text{max}/ne$, with $n=1$ (blue), 2 (green), 4 (red), 6 (cyan), 8 (magenta), 10 (yellow), and 12 (black). As we lower the threshold, the predictivity decreases as the bounded region increases. }
\end{center}
\end{figure}

Predictive models have a sharp prior predictive distribution, and therefore we do not expect a considerable change in their predictivities as we change the threshold. An example of this kind of models is RCHIo, which is introduced in Appendix \ref{App:PredInflModels}, and has the highest predictivity among the 85 models considered in this paper. Fig.~\ref{fig:RCHIoContours} shows the contours bounding regions with an evidence above $\bar{\mathcal{E}}=\mathcal{E}_\text{max}/ne$, with $n=1$ (blue), 2 (green), 4 (red), 6 (cyan), 8 (magenta), 10 (yellow), and 12 (black). The predictivity of RCHIo ranges between $P_r=0.982$ ($n=1$) and $P_r=0.974$ ($n=12$), which corresponds to a change of $0.8\%$.

It is important to notice that lowering the threshold always generates a lower predictivity as it increases the high evidence region (bounded region in Fig.~\ref{fig:RCHIoContours}). In this sense the value of predictivity itself can be quite arbitrary, although it has a particular meaning as $\%$ of population with evidence higher than the threshold. However, when we lower the threshold for all models, we can see that the high evidence region of models with a spread prior predictive distribution will change more than those with a sharp distribution. 

\begin{figure}[h]
\begin{center}
\scalebox{0.4}{\includegraphics{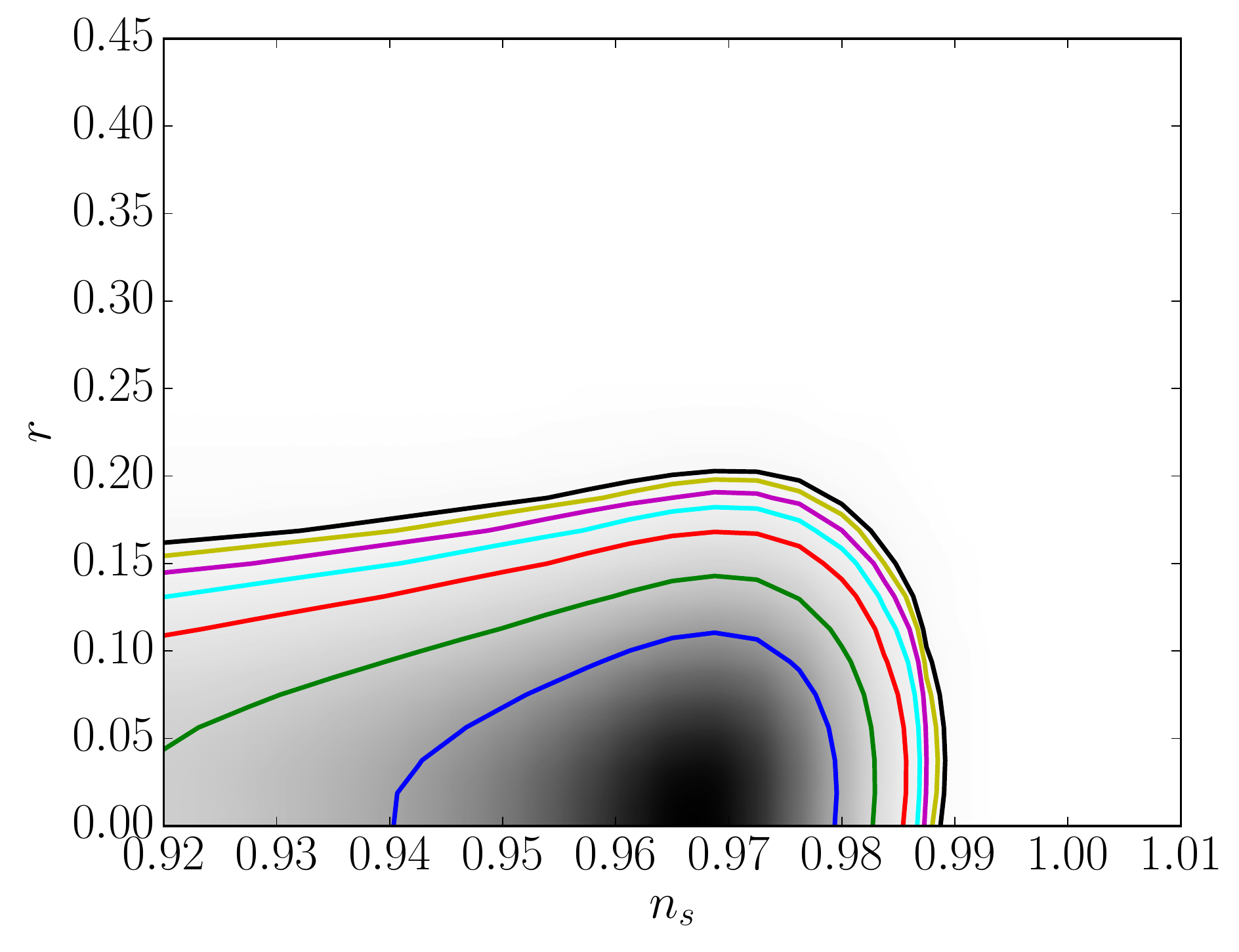}}
\caption{\label{fig:HIContours} Prior predictive distribution for HI. Darker regions have higher evidence. We also show the contours bounding regions with evidence higher than $\mathcal{E}_\text{max}/ne$, with $n=1$ (blue), 2 (green), 4 (red), 6 (cyan), 8 (magenta), 10 (yellow), and 12 (black). As we lower the threshold, the predictivity decreases.}
\end{center}
\end{figure}

To show this point we analyse the predictivity for HI for different thresholds. Fig.~\ref{fig:HIContours} shows the contours bounding regions with an evidence above $\bar{\mathcal{E}}=\mathcal{E}_\text{max}/ne$, with $n=1$ (blue), 2 (green), 4 (red), 6 (cyan), 8 (magenta), 10 (yellow), and 12 (black). The predictivity of HI ranges between $P_r=0.923$ ($n=1$) and $P_r=0.701$ ($n=12$), which corresponds to a change of $24\%$. As expected, there is a bigger change in the predictivity of HI than for RCHIo, as from the plots we could see that the prior predictive distribution of HI is more spread, as it covers a larger grey area.

A sensible choice of threshold would be one that is able to take into account the spread of the distributions. As we can see from Fig.~\ref{fig:RCHIoContours} and \ref{fig:HIContours}, as we lower the threshold we take into account the spread of the distribution and the high evidence regions can grow considerably, but the contours start becoming closer and closer, as the distributions start decaying faster. At some point, the contours stabilise and there is no relevant change in the predictivity if we lower the threshold. From this point of view, with these two models as the only information, a good choice for the threshold could be $\bar{\mathcal{E}}=\mathcal{E}_\text{max}/6e$ (cyan curve).

\begin{figure}[h]
\begin{center}
\scalebox{0.4}{\includegraphics{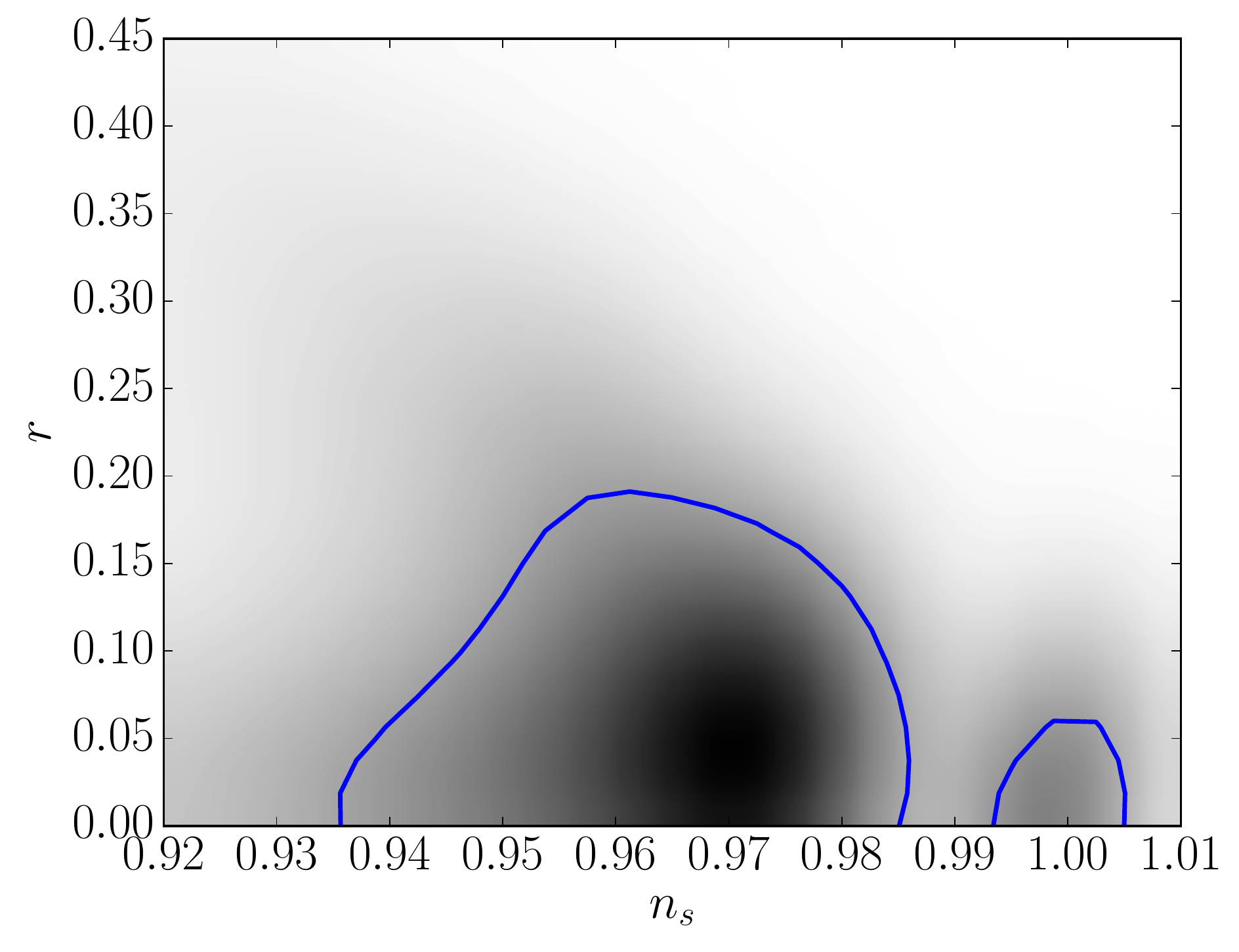}}
\scalebox{0.4}{\includegraphics{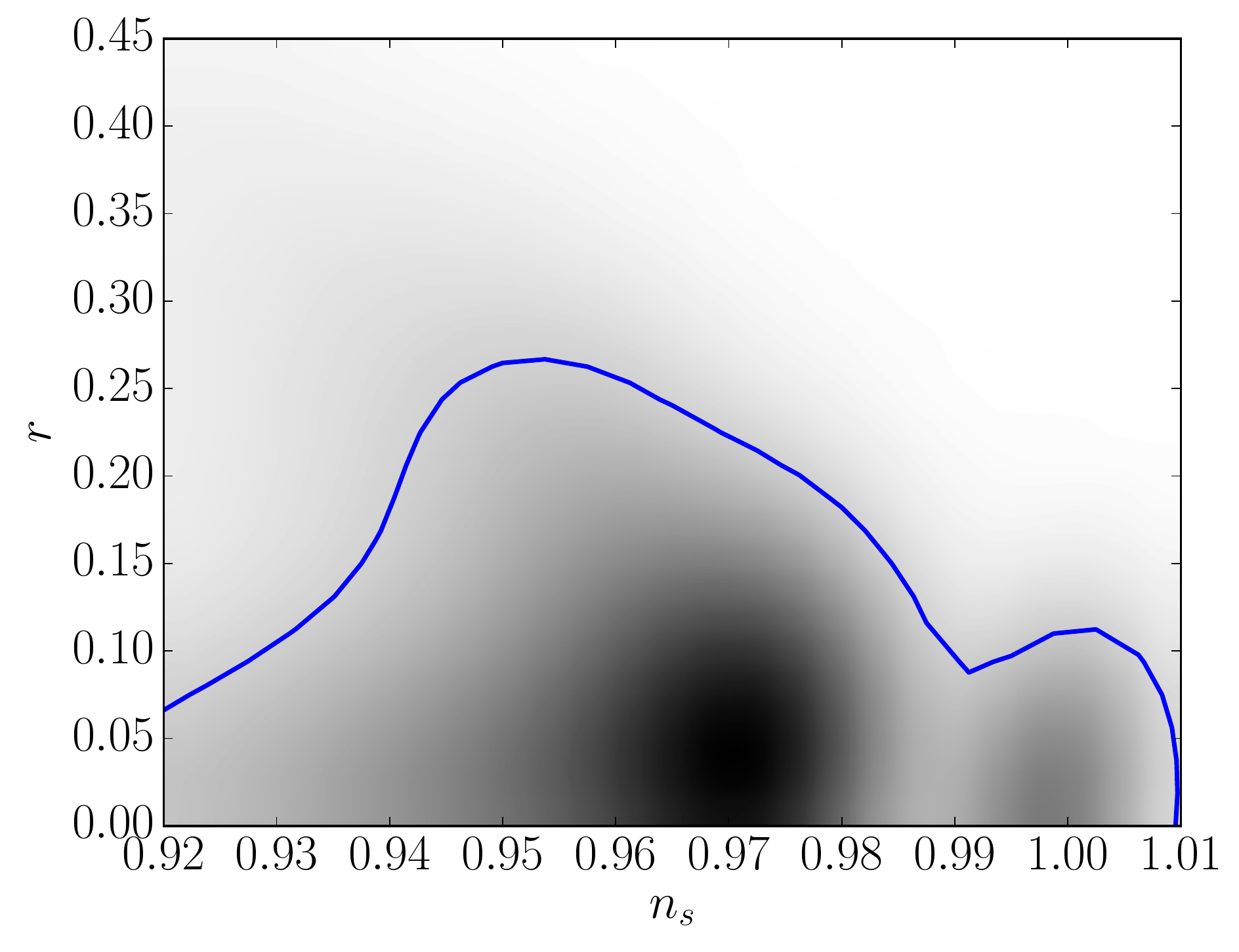}}
\caption{\label{fig:InflContour12} Prior predictive distribution of the inflationary paradigm. Darker regions have higher evidence. We also show the contours bounding regions with evidence higher than $\mathcal{E}_\text{max}/ne$, with $n=1$ (top) and $n=2$ (bottom).}
\end{center}
\end{figure}

\FloatBarrier

However, other factors can come also into play when deciding the threshold value. Particular attention is needed if a prior predictive distribution has many peaks, as a high threshold might leave out some peaks of the distribution. An example of this is the inflationary paradigm. Fig.~\ref{fig:InflContour12} shows the prior predictive distribution of the inflationary paradigm for thresholds $\mathcal{E}_\text{max}/ne$, with $n=1$ (top) and $n=2$ (bottom).

As we can see in Fig.~\ref{fig:InflContour12}, when we set $n=1$ (what we used for the toys models) we leave outside two peaks around $(n_\text{s}=0.93, r=0)$ and $(n_\text{s}=0.95, r=0.2)$. For this reason, when lowering the threshold to $n=2$ a big change occurs. The predictivity goes from $P_r=0.818$ to $P_r=0.608$. Again, a sensible choice of threshold would be one where the contours stabilise and get close enough to each other to not generate any considerable change in the predictivity. This will happen when all the peaks are included.

\begin{figure}[h]
\begin{center}
\scalebox{0.4}{\includegraphics{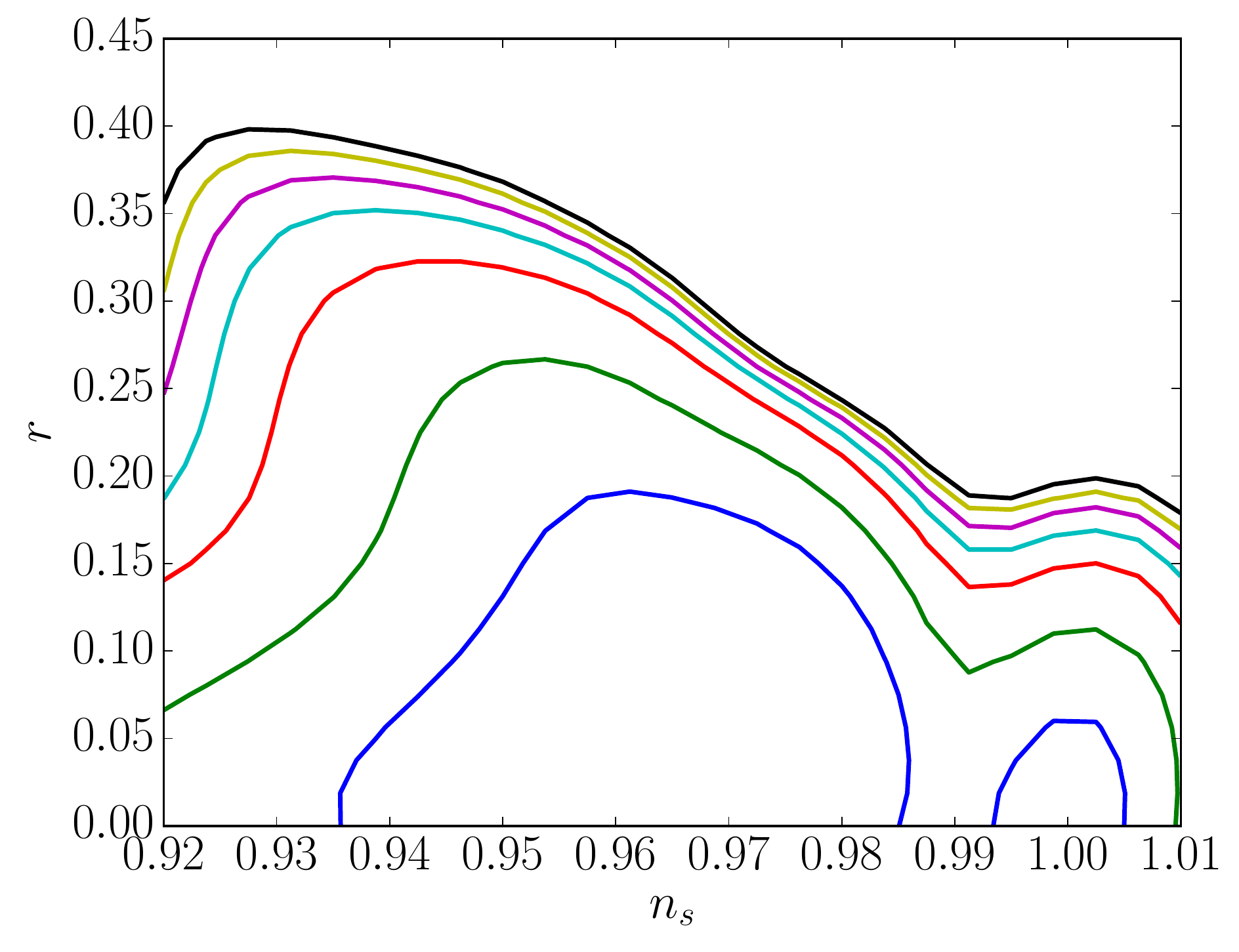}}
\caption{\label{fig:InflContours} Contours for the inflationary paradigm bounding regions with evidence higher than $\mathcal{E}_\text{max}/ne$, with $n=1$ (blue), 2 (green), 4 (red), 6 (cyan), 8 (magenta), 10 (yellow), and 12 (black). As we lower the threshold, the predictivity increases.}
\end{center}
\end{figure}

Fig.~\ref{fig:InflContours} shows the contours for the inflationary paradigm bounding regions with an evidence above $\bar{\mathcal{E}}=\mathcal{E}_\text{max}/ne$, with $n=1$ (blue), 2 (green), 4 (red), 6 (cyan), 8 (magenta), 10 (yellow), and 12 (black). In this plot we do not show the prior predictive distribution but only the contours as the former one changes with different thresholds because the prior of each model changes. We can see that the contours stabilise from $\bar{\mathcal{E}}=\mathcal{E}_\text{max}/8e$ (magenta curve). For this reason we finally choose this threshold to use throughout the paper. Notice that with this value of threshold, according to the Jeffreys' scale, a high evidence region is one that includes only events that are inconclusive or weakly disfavoured compared to event with the maximum evidence $\mathcal{E}_\text{max}$.

\FloatBarrier

\section{Experimental noise}\label{App:ExNoise}

The value of predictivity is contingent on the experimental noise. That is clear from the definition in Eq.~(\ref{Pred}), as the evidence depends on the experimental noise. In this section we would like to quantify the change on the predictivities of some inflationary models as the noise changes.

\begin{figure}[h!]
\begin{center}
\scalebox{0.4}{\includegraphics{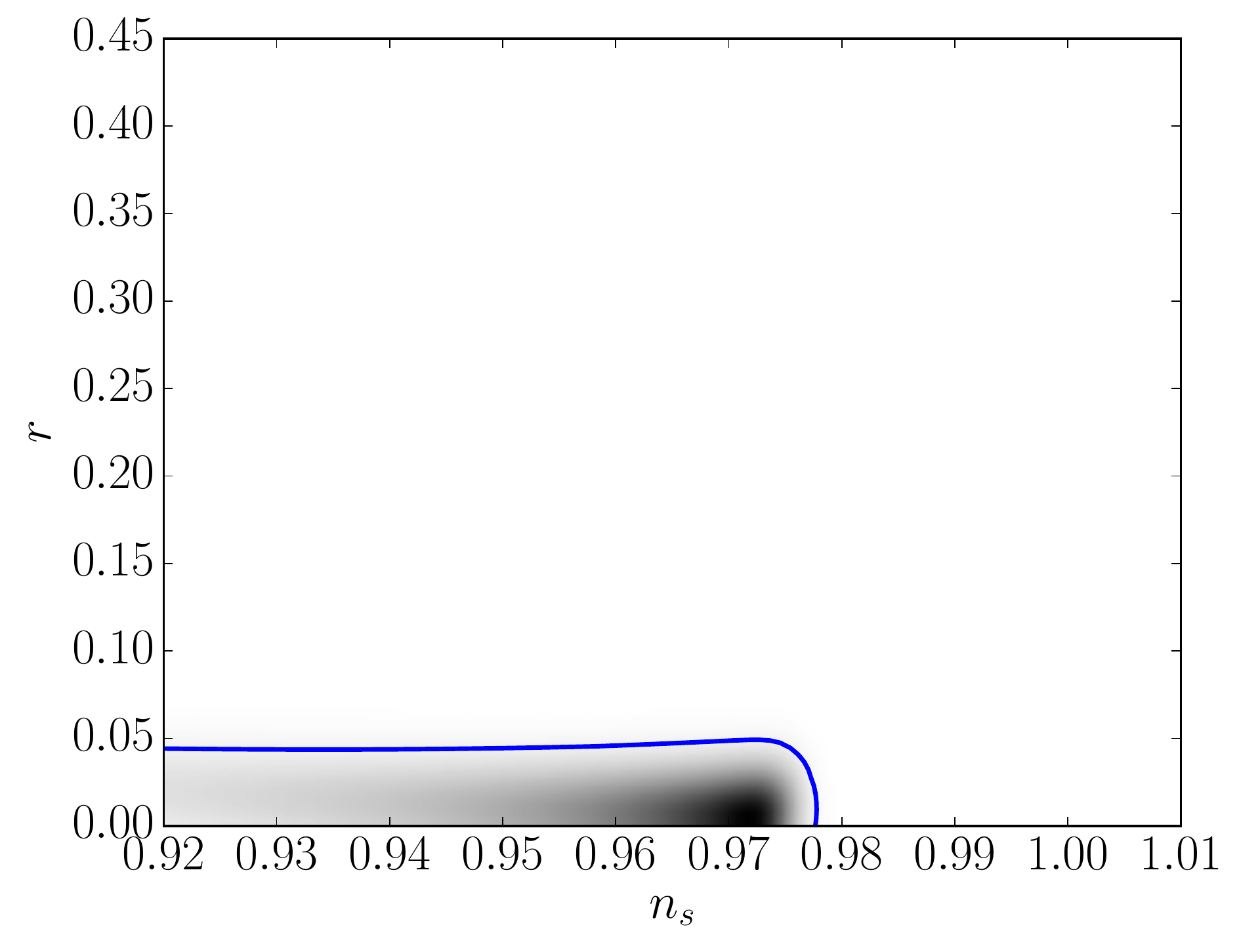}}
\caption{\label{fig:HInoise4} Prior predictive distribution for HI and contour bounding region with evidence higher than $\mathcal{E}_\text{max}/8e$ (blue), when experimental noise is reduced to a quarter of the best-fit Gaussian Planck noise. Darker regions have higher evidence. The predictivity in this case is $P_r=0.935$.}
\end{center}
\end{figure}

For HI we show the prior predictive distribution and the high evidence contour in Fig.~\ref{fig:HInoise4}, when the experimental noise is reduced to a quarter of the best-fit Gaussian Planck noise showed in Eq.~(\ref{CorrMatrix}). The predictivity of HI increases from $P_r=0.726$ (with Planck noise) to $P_r=0.935$ (with a quarter of Planck noise), which corresponds to a $29\%$ of change. 

\begin{figure}[h!]
\begin{center}
\scalebox{0.4}{\includegraphics{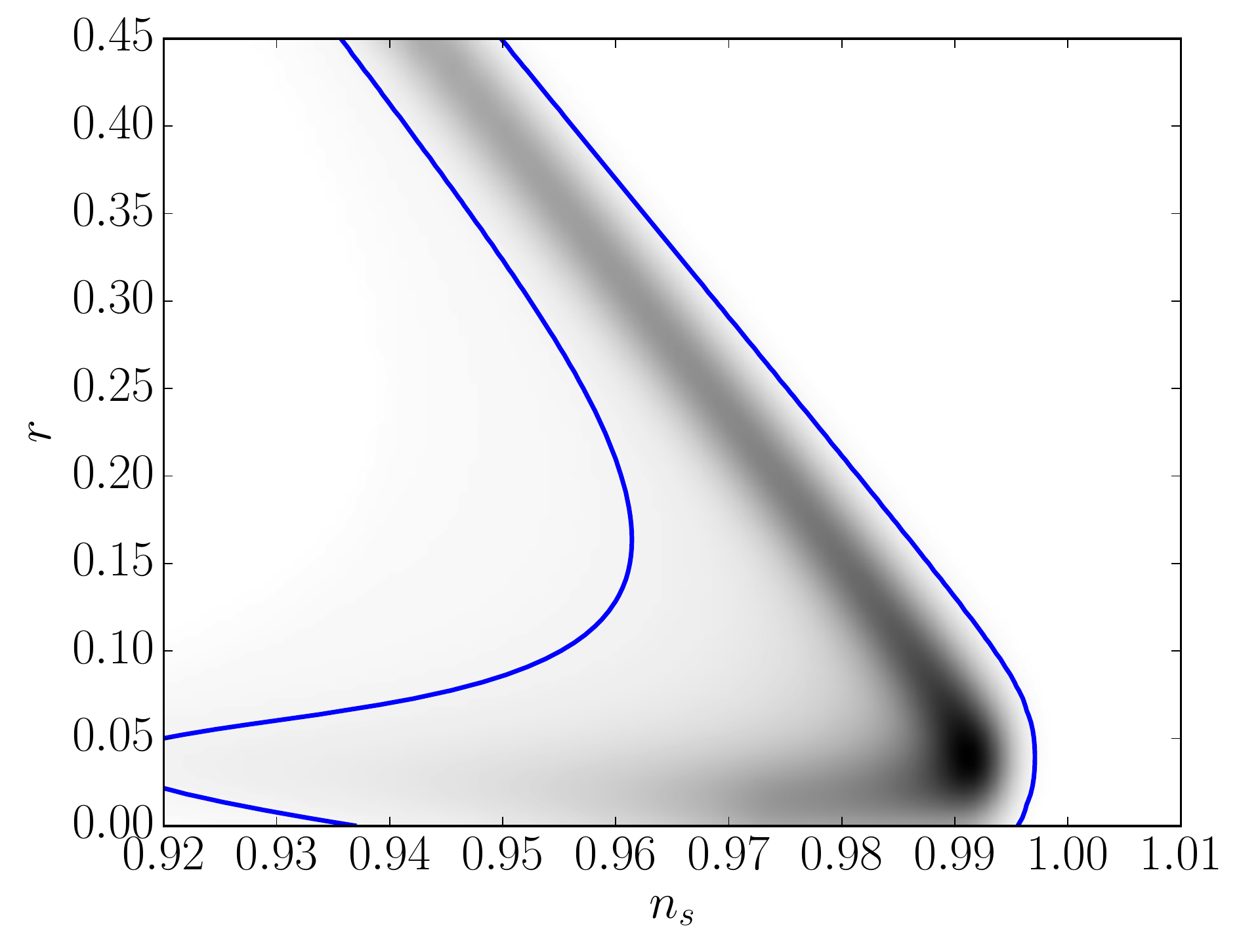}}
\caption{\label{fig:RPI3noise4} Prior predictive distribution for RPI3 and contour bounding region with evidence higher than $\mathcal{E}_\text{max}/8e$ (blue), when experimental noise is reduced to a quarter of the best-fit Gaussian Planck noise. Darker regions have higher evidence. The predictivity in this case is $P_r=0.676$.}
\end{center}
\end{figure}

For more unpredictive models such as RPI3, we find that the predictivity increases from $P_r=0.285$ (with Planck noise) to $P_r=0.676$ (with a quarter of Planck noise), which corresponds to a $137\%$ of change. Fig.~\ref{fig:RPI3noise4} shows the prior predictive distribution of RPI3 and the high evidence contour, when the experimental noise is reduced to a quarter of the best-fit Gaussian Planck noise given by Eq.~(\ref{CorrMatrix}).

As we can see with these two examples, as we lower the noise, the predictivities improve, and the change of predictivities can be considerably large. Due to the dependence of the predictivities on the experimental noise, it is important to clarify that the results of this paper estimate the testability of models or paradigms today, i.e.~with measurements of $r$ and $n_\text{s}$ with a Planck-like experiment. However, in the future, more accurate or new measurements would improve all the predictivities.


\bibliographystyle{JHEP}
\bibliography{refs-macarena}

\end{document}